\documentclass{agujournal2018}
\usepackage{apacite}
\usepackage{url} 

\draftfalse

\journalname{arXiv}

\begin{document}

%
%


\title{Variations in the Ionospheric Peak Altitude at Mars in Response to Dust Storms: 13 Years of Observations from the Mars Express Radar Sounder}

%
%




\authors{Z. Girazian\affil{1}, Z. Luppen\affil{1}, D. D. Morgan\affil{1}, F. Chu\affil{1}, L. Montabone\affil{2}, E. M. B. Thiemann\affil{3}, D. A. Gurnett\affil{1}, J. Halekas\affil{1}, A. J. Kopf\affil{1}, F. N\v{e}mec\affil{4}}

 \affiliation{1}{Department of Physics and Astronomy, University of Iowa, Iowa City, IA, USA}
 \affiliation{2}{Space Science Institute, Boulder, CO, USA}
 \affiliation{3}{Laboratory for Atmospheric and Space Physics, University of Colorado Boulder, Boulder, CO, USA}
 \affiliation{4}{Faculty of Mathematics and Physics Charles University, Prague, Czech Republic}





\correspondingauthor{Zach Girazian}{zachary-girazian@uiowa.edu}




\begin{keypoints}

\item We show how the ionospheric peak altitude at Mars varies during six different dust storms.

\item The peak altitude rises 10-15 km during each dust storm.

\item Dust storms increase the peak altitude's variability, suggesting they enhance dynamical processes coupling the lower and upper atmospheres.

\end{keypoints}

%
%


\begin{abstract}
Previous observations have shown that, during Martian dust storms, the peak of the ionosphere rises in altitude. Observational studies of this type, however, have been extremely limited. Using 13 years of ionospheric peak altitude data from the Mars Advanced Radar for Subsurface and Ionosphere Sounding (MARSIS) instrument on Mars Express, we study how the peak altitude responded to dust storms during six different Mars Years (MY). The peak altitude increased $\sim$10-15 km during all six events, which include a local dust storm (MY 33), three regional regional dust storms (MYs 27, 29, and 32), and two global dust storms (MYs 28 and 34). The peak altitude's orbit-to-orbit variability was exceptionally large at the apexes of the MY 29 and MY 32 dust seasons, and dramatically increased during the MY 28 and MY 34 global dust storms. We conclude that dust storms significantly increase upper atmospheric variability, which suggests that they enhance dynamical processes that couple the lower and upper atmospheres, such as upward propagating gravity waves or atmospheric tides. 

\end{abstract}

%
%

%


%
%
%
%

    \section{Introduction}
    \label{sec:introduction}
    
	Dust storms have proven to be an important source of atmospheric variability at Mars. Dust particles lifted into the atmosphere are heated by solar radiation, which causes the atmosphere to expand, and global circulation patterns to change \citep{haberle1993,bougher1997,heavens2011,wolkenberg2018}. Dust also affects the distribution of water vapor, which has consequences for atmospheric photochemistry, hydrogen escape, and climate evolution \citep{chaffin2017,heavens2018,daerden2019,krasnopolsky2019,vandaele2019}. Although the dust itself is mostly confined to altitudes below 80 km \citep{clancy2010}, the effects of the dust extend well into the thermosphere ($>$100 km) - even to geographic locations that are far from where the dust originated \citep{withers2013a,liu2018}.  
	
	The response of the thermosphere to lower atmospheric dust is generally marked by a rapid increase in the neutral density at a fixed altitude, followed by a slow density decay back to nominal levels over several weeks \citep{keating1998,lillis2010b,england2012,withers2013a,zurek2017}. During typical dust events, thermospheric neutral densities increase by a factor of $\sim$1.5-3.0 at fixed altitudes \citep{withers2013a,zou2016,liu2018}, but thermospheric neutral temperatures do not drastically change at fixed pressure levels \citep{mcelroy1977,wang2003b,fang2019a}. When thermospheric pressure surfaces rise in response to dust loading, the peak of the ionosphere - which typically forms between 120-150 km and at a fixed pressure level \citep{withers2009a} - rises in altitude. 
	
	Elevated ionospheric peak altitudes during dust storms have been observed by radio occultation (RO) experiments on several spacecraft, the first being Mariner 9, which arrived at Mars in 1971 during a global event. Mariner 9 observed that, during the global dust storm, the ionospheric peak altitude was $\sim$20 km higher than usual \citep{hantsch1990}, and then slowly decayed back to typical values during the dust storm's waning stages \citep{withers2013a}. The RO experiments on the Mars Global Surveyor (MGS) and Mars Atmosphere and Volatile EvolutioN (MAVEN) spacecraft have also observed the ionospheric peak rise during regional dust storms. MGS observed the peak rise $\sim$5 km during a dust storm at Solar Longitude (L$_s$) 130$^{\circ}$ in Mars Year (MY) 27 \citep{withers2013a,qin2019}, and MAVEN observed the peak rise $\sim$10 km during a dust storm at L$_s$ $\sim$305$^{\circ}$ in MY 33 \citep{withers2018}.
    
    Motivated by the limited amount of observations showing how the ionospheric peak responds to dust storms, we utilize a 13-year span of peak altitude measurements from the Mars Advanced Radar for Subsurface and Ionosphere Sounding (MARSIS) instrument on Mars Express (MEX) \citep{picardi2004,gurnett2005}. By combining these data with dust optical depth measurements from the same time period \citep{montabone2015}, we investigate how dust affected the ionospheric peak altitude during six different MYs. In two of these years, MY 28 and MY 34, there was a global dust storm. Our objectives are (1) to compare the response of the peak altitude during six different dust storms and (2) to determine if dust storms affect the variability of the peak altitude. 
 
    The paper is organized as follows. In Section~\ref{sec:theory}, we present the theory that describes the formation of the ionospheric peak, and explain how variations in the ionospheric peak altitude can be used to estimate changes in the thermospheric pressure. In Section~\ref{sec:data}, we describe the data sets that are used in our analysis. In Section~\ref{sec:dustseasons}, we show how dust affected the peak altitude during several MYs. In Section~\ref{sec:conclusions}, we discuss our results and present our conclusions.

    \section{Theory of the Ionospheric Peak Altitude}
    \label{sec:theory} 
    
     The main peak of the Martian ionosphere is well-described by Chapman theory \citep{chapman1931a,schunk2009,withers2009a,girazian2013,mendillo2017}. Chapman theory predicts that, under photochemical equilibrium conditions, the ionospheric peak forms at the altitude where the optical depth of ionizing extreme ultraviolet (EUV) photons is equal to one \citep{breus2004,withers2009a}. Mathematically, this is approximated by
     \begin{equation}
     \label{eq1}
     n(h_{max}) \sigma H \sec{\chi} = 1
     \end{equation}
    
    \noindent{where} $n(h_{max})$ is the neutral CO$_2$ density at the peak altitude $h_{max}$, $\sigma$ is the CO$_2$ absorption cross section at EUV wavelengths, $H$ is the neutral scale height, and $\chi$ is solar zenith angle (SZA). Equation~\ref{eq1} requires many simplifying assumptions to be valid \citep{schunk2009,withers2009a} but has proven to be an adequate description of the ionospheric peak at Mars \citep{withers2009a,girazian2013,fallows2015,mendillo2017}. Using Equation~\ref{eq1}, several studies have shown that observed changes in the ionospheric peak altitude can be used to estimate variations in the thermospheric neutral density or pressure \citep{bougher2001,bougher2004,withers2013a,zou2011,zou2016,qin2019}.
    
    For our purposes, we are interested in using Equation~\ref{eq1} to quantify how the peak altitude rises or falls in response to changes in the thermospheric pressure. Equation~\ref{eq1} is derived under the assumption of a static, isothermal atmosphere so that it can be recast as
    
    \begin{equation}
    \label{eq2}
    n_{0}e^{-(h_{max}-h_{0})/H} \sigma H \sec{\chi} = 1.0
    \end{equation}
    
    \noindent{where} $n_{0}$ is the neutral density at some reference altitude $h_{0}$. If $n_0$ or $H$ changes, $h_{max}$ must rise or fall such that Equation~\ref{eq2} remains satisfied:
    
    \begin{equation}
    \label{eq3}
    n_{0_1}e^{-(h_{max_{i}}-h_{0})/H_i} H_i = n_{0_f}e^{-(h_{max_f}-h_{0})/H_f} H_f =  1.0
    \end{equation}
    
\begin{figure}[ht!]
\centering
\includegraphics[width=1.0\textwidth]{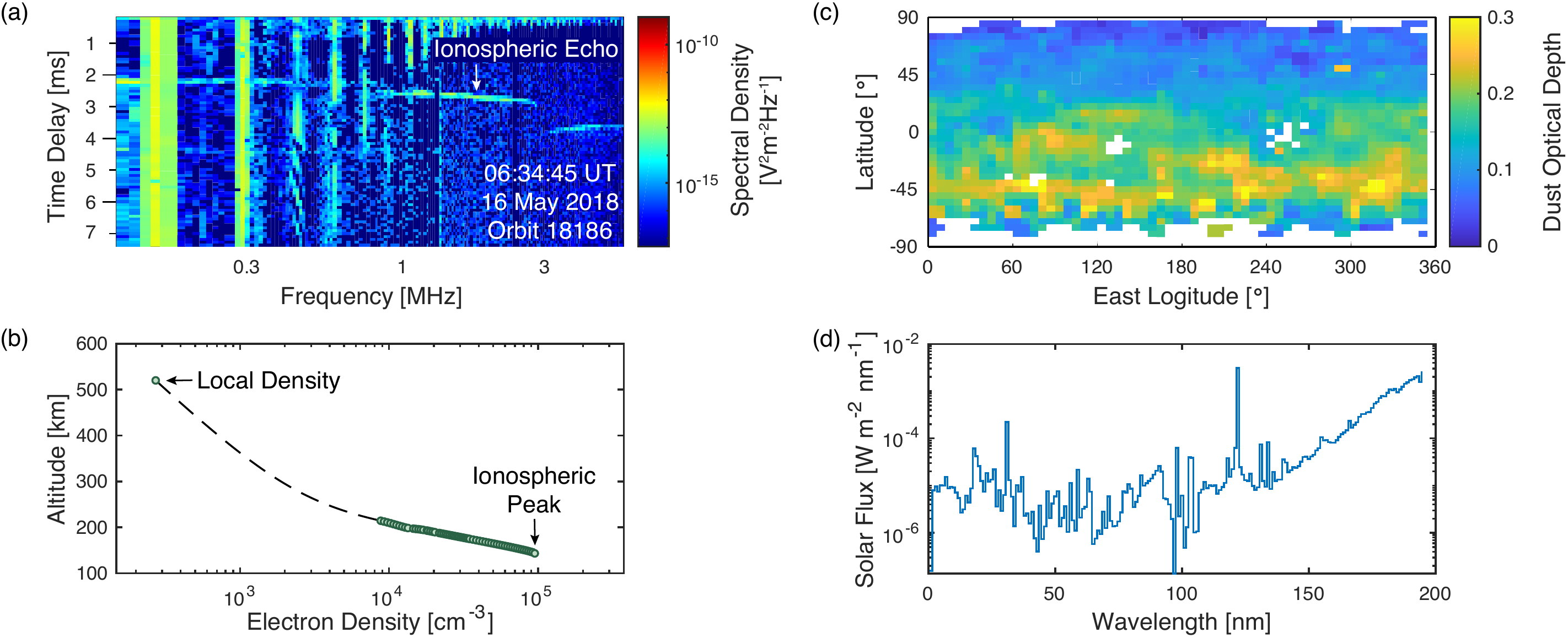}
\caption{Examples of the four data products used in our study, all of which are from 16 May 2018. \textbf{a)} MARSIS ionogram showing the vertical stripes at low frequencies used to derive the local electron density, and the ionospheric echo that captures the electron density down to the ionospheric peak. \textbf{b)} The electron density profile derived from the MARSIS ionogram shown in Panel a. The dashed line marks the measurement gap as explained in the text. \textbf{c)} The dust optical depth map from \citet{montabone2015}. \textbf{d)} The solar EUV spectrum from the FISM-M model \citep{thiemann2017}. 
}
\label{examples}
\end{figure}

   \noindent{where the subscripts $i$ and $f$ represent the initial and final states, respectively. In Equation~\ref{eq3}, we have removed the SZA dependence by assuming that it is fixed. Next, by setting the reference altitude to $h_{0} = h_{max_i}$, Equation~\ref{eq3} becomes
   
    \begin{equation}
    \label{eq4}
      \frac{n_{0_f} H_f}{n_{0_i} H_i} = \exp\left(\frac{\Delta h_{max}}{H_f}\right)
    \end{equation}
    
     \noindent{where} $ \Delta h_{max} = h_{max_f} - h_{max_i}$. According to Equation~\ref{eq4}, if we assume that $H_i = H_f$, then a factor of $\sim$3 increase in $n_0$ results in the peak altitude increasing by one scale height. Furthermore, Equation~\ref{eq4} states that increases in the neutral density or scale height during dust storm onset will cause the peak altitude to rise, while decreases in the density or scale height during the waning stages of a dust storm will cause the peak altitude to fall. 
     
     Equation~\ref{eq4} can also be recast in terms of atmospheric pressure. The pressure is given by $P = \rho g H$, where $P$ is pressure, $\rho$ is mass density, and $g$ is the gravitational acceleration, which is assumed to be constant. This equation for atmospheric pressure is equivalent to the ideal gas law for an isothermal atmosphere with a fixed scale height $H$. The assumption of a fixed $H$ is used throughout this work because observations and modeling suggest that dust storms significantly affect thermospheric densities, but only modestly affect thermospheric temperatures \citep{mcelroy1977,wang2003b,liu2018,fang2019a,qin2019}. This assumption does, however, add uncertainty to our analysis.
     
     Substituting the pressure into Equation~\ref{eq4} gives 
     
    \begin{equation} 
    \label{eq5}
    \frac{P_f}{P_i} = \exp\left(\frac{\Delta h_{max}}{H}\right).
    \end{equation}
    
    \noindent{Here}, $P_i$ and P$_f$ are the atmospheric pressures at the ionospheric peak and $H$ has been assumed to not change. Equation~\ref{eq5} states that, at a fixed SZA, the peak altitude forms at a constant atmospheric pressure level. Equation~\ref{eq5} was used in the study by \citet{withers2013a} to estimate changes in the thermospheric pressure during the waning stages of the Mariner 9 dust storm.
   
    In addition to allowing one to quantify how the peak altitude responds to changes in the thermospheric density, scale height, or pressure, Eqs.~\ref{eq1}, \ref{eq4}, and \ref{eq5} summarize the conditions that control variations in the ionospheric peak altitude. Equation~\ref{eq1} can be inverted to show that the peak altitude increases with increasing SZA proportional to $ln(\sec\chi)$, indicating a steep rise in the peak altitude near the day-night terminator \citep{withers2009a,fallows2015}. Equation~\ref{eq5} describes how the peak altitude can vary with latitude and local time due to diurnal pressure gradients in the thermosphere \citep{zou2011,bougher2015}, and further shows that any physical process that alters the thermospheric pressure will also alter the peak altitude. These processes include, but are not limited to (1) the annual variation in solar irradiance at Mars due to its large orbital eccentricity; (2) atmospheric tides and waves; (3) solar extreme ultraviolet (EUV) heating; and (4) atmospheric circulation patterns. All of these processes must be considered when attempting to identify variations in the peak altitude caused solely by dust storms.

    
    \section{Data}
    \label{sec:data}
    
    We use three types of data: ionospheric peak altitudes from the MARSIS radar sounder, dust optical depths compiled from several instruments, and solar EUV irradiance from an empirical model. In this section, we describe each data set and our processing techniques.
    
    \subsection{Peak Altitudes}
    
    Ionospheric peak altitudes are derived from MARSIS radar sounding observations. MARSIS sounds the ionosphere during the periapsis segment of the $\sim$7-hour, near-polar orbit of MEX. When MARSIS sounds the ionosphere, it transmits radio pulses and records the return echoes from pulses that are reflected off the ionosphere. The transmitter sweeps through 160 qausi-logarithmically spaced frequencies between 0.1 and 5.4 MHz over 1.257 seconds. Each sweep produces an ionogram - the echo intensity as a function of frequency and time delay. MARSIS makes frequency sweeps every 7.54 seconds while MEX is below $\sim$1500 km, returning several hundred ionograms during each periapsis pass. 
    
    Figure~\ref{examples}a shows an ionogram from 16 May 2018. The vertical stripes at low frequencies are a common feature; they are produced by electron plasma oscillations induced by the radio transmitter during ionospheric soundings. The spacings between the stripes are used to determine the local electron density at the spacecraft location \citep{duru2008,gurnett2005,gurnett2008}. The thin horizontal stripe between $\sim$1.0-3.0 MHz is the ionospheric echo, which represents radar pulses that were reflected off the ionosphere below $\sim$200 km \citep{gurnett2005,gurnett2008}. The highest frequency of the ionospheric echo, near 3 MHz in this example, is the reflection from the ionospheric peak. The time delay at this frequency in the ionospheric trace is related to the altitude of the peak. The horizontal stripe at frequencies greater than 3 MHz, with a time delay of $\sim$4 ms, is the return echo from the surface of Mars. 
    
    If an ionogram provides a local electron density measurement, and also has a clear ionospheric trace, then it can be inverted into an altitude profile of the electron density. The inversion procedure requires one to make an assumption about the shape of the electron density profile between the altitude of MEX, where the local density is measured, and the altitude of the first electron density measurement in the ionospheric trace. In this work, we adopt the inversion technique described in \citet{nemec2016}, which assumes that the shape of the density profile within the measurement gap is characterized by a lower and upper topside scale height, with a smooth transition between them. Figure~\ref{examples}b shows the electron density profile that was derived by applying this technique to the ionogram shown in Figure~\ref{examples}a. Once inverted into an electron density profile, the peak altitude is easily extracted as marked in Figure~\ref{examples}b.

    
    We use MARSIS ionograms obtained between 11 July 2005 and 14 July 2018. All of the ionograms obtained up to 22 May 2016 were inverted into electron density profiles. Only a subset of the ionograms obtained after 22 May 2016 were used, because they have not yet been inverted into electron density profiles due to the time consuming and hands-on processing techniques that are required \citep{gurnett2008,morgan2008}. In lieu of this, we have inverted two subsets of ionograms from after this date specifically for this study. The first subset of ionograms is from the month of Jan. 2017. This month was targeted because elevated peak altitudes were observed by the MAVEN RO experiment during a dust storm \citep{withers2018}. The second subset is from 14 May 2018 through 14 July 2018. This period was targeted because it covers a significant portion of the 2018 global dust storm \citep{guzewich2019,vandaele2019}.   

    After inverting the ionograms into electron density profiles and extracting their peak altitudes, we filter the data set based on several criteria. First, we keep density profiles only if they monotonically decrease with increasing altitude, which is a requirement of the \citet{nemec2016} inversion technique. Second, we remove any peak altitudes below 80 km or above 220 km, well outside the expected peak altitude and likely the result of bad inversions \citep{withers2009a,fox2012,nemec2016,vogt2017}. Third, we limit the profiles to times when MEX was below 1000 km. After applying these criteria, the complete data set includes more than 180,000 peak altitude measurements from 2401 MEX orbits.

    Uncertainties in the peak altitudes are, at best, $\pm$7 km, as determined by the intrinsic 91.4 $\mu$s time resolution of the MARSIS receiver. The \citet{nemec2016} inversion technique also adds to this uncertainty, given that it relies on an assumption about the shape of the electron density profile within the measurement gap. In Section~\ref{sec:dustseasons}, we analyze orbit-averaged peak altitudes (from a limited SZA range), and assign uncertainties based strictly on the spread of the observed peak altitudes during each orbit. In particular, for each orbit, we define the uncertainty in the orbit-averaged peak altitude as the standard deviation of the peak altitudes used to calculate the average.

    \subsection{Dust Optical Depths}
    \label{sec:dustdata}
    
     Dust optical depth maps are derived by combining measurements from several instruments, as described in \citet{montabone2015}. The maps provide a continuous measure of the dust content in the lower atmosphere, with good coverage in latitude, longitude, and L$_s$ from 1999 until present. The dust data for the time period considered here are synthesized from infrared observations by the Mars Odyssey Thermal Emission Imaging System (THEMIS) \citep{christensen2004} and the Mars Reconnaissance Orbiter Mars Climate Sounder (MCS) \citep{mccleese2007}. The maps provide the optical depth at 9.3 $\mu$m absorption, normalized to an atmospheric pressure level of 610 Pascals. An example of a dust optical depth map is shown in Figure~\ref{examples}c.  
     
     Throughout this work, we use the observation-only dust maps that sometimes have incomplete coverage in latitude and longitude \citep{montabone2015}. We also use dust maps that were developed specifically for the MY 34 global dust storm and this special issue. These MY 34 dust maps use estimated column dust opacities from MCS as described in detail in \citet{montabone2019}. We use the v2.5 version of the maps.
     
    
     We assign a ``local'' average and global average dust optical depth to each MARSIS electron density profile. The local average accounts for our expectation that the ionospheric peak will respond to dust storms that are nearby. The global average accounts for our expectation that the ionospheric peak might also respond to dust storms that are far away, due to their effects on atmospheric circulation \citep{bell2007,withers2013a,withers2018}. To assign the global average, we match each MARSIS electron density profile with the dust optical depth map from the closest date, and then average the dust optical depth map over all latitudes and longitudes. Since the dust maps have a resolution of $\sim1^{\circ}$ in L$_s$, the global average dust optical depth is constant during each MEX orbit. 
     We assign the local average in a similar way but only after restricting the dust data to the latitude range of the MARSIS observations during each orbit.

    
    
    \subsection{Solar EUV Irradiances}
    
     Solar EUV irradiances are from the Flare Irradiance Spectral Model-Mars (FISM-M), which is an empirical model that provides daily-averaged solar EUV spectra at Mars \citep{thiemann2017}. The spectra have 1 nm resolution and cover wavelengths between 0.5-189.5 nm. An example FISM-M spectrum, from 16 May 2018, is shown in Figure~\ref{examples}d.   
     
    
     We assign a single number of the EUV irradiance (W m$^{-2}$) to each MARSIS electron density profile. We do this by first matching each electron density profile with the solar EUV spectrum from the closest date, then integrating the matched EUV spectrum over wavelengths between 0.5-92.5 nm. The cutoff wavelength of 92 nm is chosen because it is the longest wavelength photon that can ionize O and CO$_2$, which are the most abundant neutral species in the thermosphere of Mars \citep{schunk2009,girazian2013,girazian2015b,mahaffy2015a}. 
    
    \subsection{Overview of the Data}
    \label{sec:overview}
      
    Figures~\ref{overview}a-c show the 13-year time series of ionospheric peak altitudes along with the SZAs and geographic latitudes of the MARSIS observations. Data gaps are present throughout the time series because MARSIS frequently toggles between ionospheric and subsurface sounding modes, and does not make observations during eclipse seasons when there is insufficient sunlight to recharge the MEX solar panels. The error bars in Panels b-c show the $\sim$80$^{\circ}$ in latitude and the $\sim$30$^{\circ}$} in SZA that MARSIS covers during a typical periapsis pass. We also note that, from one periapsis pass to the next, the observational SZA and latitude are nearly identical, but the longitude is shifted by $\sim$100$^{\circ}$.
    
    Figure~\ref{overview}d shows the solar EUV irradiance and the inverse-square of the Mars-Sun distance during the MARSIS observations. The latter is representative of the solar insolation at Mars, which varies annually due to the planet's eccentric orbit around the Sun. The 13-year period covers more than a complete solar cycle, starting with the declining phase of Solar Cycle 23 in 2005, and continuing through the declining phase of Solar Cycle 24 in 2018. The EUV irradiance varies over the 11-year solar cycle, reaching a minimum in 2008, a maximum in 2014, and also annually due to the varying Mars-Sun distance.
    
    Figure~\ref{overview}e shows the global average dust optical depth from the same time period. The repeated peaks in the optical depth during each Martian Year are the signature of the well known annual dust cycle at Mars \citep{montabone2015,fang2019a}. During most years, the optical depth reaches a maximum between L$_s$ 210$^{\circ}$-240$^{\circ}$, and then exhibits a smaller, secondary peak between L$_s$ 300$^{\circ}$-340$^{\circ}$. Exceptions to the typical annual dust cycle are seen in MYs 28 and 34, when global dust storms caused large spikes in the dust optical depth at atypical L$_s$ values \citep{montabone2015,montabone2019,wolkenberg2018,guzewich2019,sanchez2019}. 
    
    Now that we have provided an overview of the data used in our study, we are ready to focus on our science objectives, which are (1) to compare the response of the peak altitude during six different dust storms and (2) to determine if dust storms affect the variability of the peak altitude. In the next section, we will focus on these objectives by closely examining how the ionospheric peak altitude responded to dust storms during six different MYs. For each MY, we analyze a subset of MARSIS peak altitudes from a limited range of L$_s$ and SZA. With the exception of MY 33, the L$_s$ range is chosen to capture the onset of a dust storm, and the SZA range is chosen to capture the smallest SZAs observed by MARSIS during that period. 
    
    To provide global context, Figure~\ref{dustmaps} shows the complete dust maps from the six MYs we perform our analysis. Each panel in Figure~\ref{dustmaps} is marked with a rectangle that shows the L$_s$ and latitude coverage of the MARSIS peak altitudes that we analyze during that MY. Figure~\ref{dustmaps} also highlights that, with the exception of MY 27, the dust maps have full latitudinal coverage during the time periods considered.
    
    The observations in MYs 27-29 provide the most favorable conditions because MARSIS observations cover dayside SZAs $<$ 55$^{\circ}$ during times when the global dust content significantly increases. The MY 29 observations at southern latitudes, and the MY 32 observations at northern latitudes, are from similar dust conditions. This allows us to compare the effects of dust in two different hemispheres. The MY 30 and MY 31 observations are not presented because dayside observations were not obtained during the onset of dusty periods. The MY 33 observations are specifically targeted to compare with MAVEN observations from the same period \citep{withers2018}. Finally, the MY 34 observations, which are from northern polar latitudes, cover the onset of the MY 34 global dust storm.
    
    Based on previous studies, we expect that high local dust content will elevate the peak altitude, and that high global dust content may elevate the peak altitude due to its effects on atmospheric circulation patterns \citep{hantsch1990,bell2007,withers2013a,zou2016,liu2018,fang2019a,qin2019}. Additionally, increases in the peak altitude might be less noticeable during low solar EUV levels when the upper atmosphere is intrinsically more variable due to increased gravity wave activity \citep{england2017,terada2017,harada2018,siddle2019}.


    \section{Peak Altitude Variations During Six Dust Storms}
    \label{sec:dustseasons}

    \subsection{Mars Year 27}
    \label{subsec:MY27}
     
     Figure~\ref{MY27} summarizes the observations from late in the MY 27 dust season between L$_s$ 305$^{\circ}$-330$^{\circ}$. In Figure~\ref{MY27}a, the MARSIS observational coverage in latitude and SZA is plotted on top of the dust optical depth map. The MARSIS data from this time period are restricted to SZAs between 20$^{\circ}$-50$^{\circ}$ to rule out SZA being a significant factor in any observed peak altitude variations. The dust map shows a rapid increase in the atmospheric dust content starting at L$_s$ 310$^{\circ}$ at most latitudes. Dust optical depth measurements at latitudes greater than $+$45$^{\circ}$ were unavailable during this period \citep{montabone2015}. Figure~\ref{MY27}b shows the global and local average dust optical depths (Section~\ref{sec:dustdata}) for each MEX orbit during which MARSIS obtained at least 10 peak altitude measurements. The optical depths increase sharply at L$_s$ $\sim$305$^{\circ}$ and then slowly decline through L$_s$ $\sim$320$^{\circ}$. 
     
     Figure~\ref{MY27}c shows the corresponding orbit-averaged ionospheric peak altitudes, with error bars representing the spread in the peak altitude measurements (1$\sigma$) from each MEX orbit. The peak altitude rapidly rises 10-15 km between L$_s$ 307$^{\circ}$-315$^{\circ}$, and then slowly descends back to pre-dust storm values over $\sim15^{\circ}$ of L$_s$. The peak altitude's rapid rise and slow descent is consistent with previous reports of the peak altitude's response to dust storms \citep{withers2013a,withers2018}. 
     
     Using the observed peak altitudes and Equation~\ref{eq5}, we can estimate how the thermospheric pressure changed during the dust storm \citep{withers2013a,qin2019}. In Equation~\ref{eq5}, we set $h_{max_1}$ to a constant reference altitude equal to the average peak altitude prior to dust storm onset (L$_s < 307^{\circ}$). Then, we set the neutral scale height, $H$, to a fixed value of 12 km, which is a typical value for the dayside thermosphere \citep{withers2006a,mahaffy2015a,zurek2017}. This choice introduces some uncertainty in the derived pressure changes because $H$ can vary between 8-16 km. With $h_{max_1}$ and $H$ fixed, Equation~\ref{eq5} gives the pressure ratio, $P_f/P_i$, at every point where there is an orbit-averaged peak altitude. This method was used by \citet{withers2013a} to estimate thermospheric pressure changes during the Mariner 9 global dust storm. We adopt their terminology by calling $P_f/P_i$ the ``relative pressure'' ($P_{rel}$). 
     
     The peak in the relative pressure, shown in Figure~\ref{MY27}d, coincides with the maximum in the dust optical depth at L$_s \sim$315$^{\circ}$, and reaches a value of $P_{rel} = 2.4 \ (\pm 2.4)$. The standard deviation of $P_{rel}$ does not take into account the range of possible $H$ values, but is large due to the highly variable peak altitudes observed during this orbit. The maximum pressure increase of 2.4 is on the same order as the value derived during the waning stages of the Mariner 9 global dust storm \citep{withers2013a}.  
     
     Next, we test if the variability of the peak altitude is affected by the dust storm. To accomplish this, we define a variability metric, $\sigma_{h_{max}}$, and compare its value before and after dust storm onset. The metric is a measure of the peak altitude's orbit-to-orbit variability. We determine $\sigma_{h_{max}}$ by first calculating the absolute differences between adjacent orbit-averaged peak altitudes, and then setting $\sigma_{h_{max}}$ equal to the average of the differences within a specified L$_s$ range. Using this procedure, we find that $\sigma_{h_{max}}$ = 4 km both before (L$_s < 312^{\circ}$) and after (L$_s > 312^{\circ}$) the peak of the storm. Thus, the dust storm did not significantly affect the variability of the peak altitude.
     
     To conclude, the peak altitude increased by 10-15 km during the MY 27 dust storm, but the variability of the peak altitude was unaffected. This MY 27 case study is summarized in Table 1, which includes a description of the MARSIS observing conditions, the dust storm, and the response of peak altitude. Table 1 also includes a summary of the five other case studies presented throughout this section.


    \subsection{Mars Year 28}
    \label{subsec:MY28}
     
    Figure~\ref{MY28} summarizes the observations from the MY 28 global dust storm between L$_s$ 250$^{\circ}$-300$^{\circ}$. During the global dust storm, the lower atmospheric dust content was atypically large, and wide-spread around the planet \citep{montabone2015,wolkenberg2018}. As Figure~\ref{MY28}a shows, the MARSIS data from this period are from SZAs 30$^{\circ}$-40$^{\circ}$, and from southern latitudes between -50$^{\circ}$ and -65$^{\circ}$. Figures~\ref{MY28}a-\ref{MY28}b show that the local and global dust optical depths begin to increase at L$_s$ $\sim$265$^{\circ}$ and reach maxima between L$_s$ $\sim$275$^{\circ}$-$\sim$280$^{\circ}$. The MARSIS observations do not cover the entire event because dust levels remained elevated through L$_s$ $\sim$320$^{\circ}$ \citep{montabone2015,wolkenberg2018}.
    

    The ionospheric peak altitude, shown in Figure~\ref{MY28}c, begins to rise at the onset of the dust storm, and continues to rise throughout the observational period. Furthermore, the variability of the peak altitude significantly increases after the peak of the dust storm (L$_s$ $\sim$280$^{\circ}$). From before dust storm onset (L$_s < 265^{\circ}$) to after the dust storm peak (L$_s > 280^{\circ}$), $\sigma_{h_{max}}$ more than doubles, increasing from 4 km to 10 km. (Although not shown, we also note that the highly variable peak altitudes have no longitudinal trend).
    
    Such variability was not observed in MY 27 (Figure~\ref{MY27}), during which the MARSIS observations covered similar latitudes and SZAs, but during which the dust optical depth was a factor of $\sim$2 smaller (Figure~\ref{MY27}). One possible explanation is that the atypically high dust levels during the MY 28 global dust storm increased upper atmospheric variability, perhaps by enhancing upward propagating waves, atmospheric circulation, or atmospheric tides \citep[and references therein]{bell2007,medvedev2013,bougher2015a}. The relative pressure, shown in Figure~\ref{MY28}d, is extremely variable after the peak of the dust storm, changing by as much as a factor of 15 across the time period, implying that the upper atmosphere is perturbed both spatially and temporally.
    
    In summary, the peak altitude increased by 10-15 km and became highly variable during the MY 28 global dust storm.

    \subsection{Mars Year 29}
    \label{sec:MY29}
    
    Figure~\ref{MY29} summarizes a small number of observations from the peak of the MY 29 dust season between L$_s$ 230$^{\circ}$-245$^{\circ}$. As Figure~\ref{MY29}a shows, the MARSIS observations during this period are from SZAs 45$^{\circ}$-55$^{\circ}$ and from southern latitudes between -65$^{\circ}$ and -75$^{\circ}$. Figures~\ref{MY28}a-\ref{MY28}b show that, at the start of observing period, the dust optical depth is already higher than usual because the MY 29 dust season began several months earlier at L$_s$ 170$^{\circ}$ (Figure~\ref{dustmaps}). Nonetheless, the local and global dust optical depths increase by 0.1 between 235$^{\circ}$-245$^{\circ}$. 
    
    The peak altitude, shown in Figure~\ref{MY29}c, potentially increases between L$_s$ 234$^{\circ}$-238$^{\circ}$, but the trend is weak owing to the peak altitude's remarkable variability. The variability metric is $\sigma_{h_{max}}$ = 11 km throughout the entire period and does not change after the global average dust optical depth increases (L$_s > 238^{\circ}$). The magnitude of the variability metric is comparable to that derived for the MY 28 global dust storm (Section~\ref{subsec:MY28}). The relative pressure, shown in Figure~\ref{MY29}d, is also remarkably variable and changes by as much as a factor of 10.
    
    Comparing the MY 29 and MY 27 (Figure~\ref{MY27}) case studies reveals some stark differences. MY 27 is marked by low variability ($\sigma_{h_{max}}$ = 4 km) and the peak altitude exhibits a clear increase in tandem with the dust optical depth. Meanwhile, MY 29 is marked by such high variability ($\sigma_{h_{max}}$ = 11 km) that the increasing peak altitude trend is comparable to the orbit-to-orbit variations.
    
    Differences in dust content during the MY 27 and MY 29 observational periods may explain these differences. The global dust optical depths in MY 27 rapidly increase by a factor of three (0.1-0.3) over 7$^{\circ}$ in L$_s$, while the optical depths in MY 29 increase by only a factor of 1.5 (0.25-0.35) over the same time period. Another consideration is that, in MY 29 the dust season was well underway when the observing period began, which may explain the exceptional variability. To contrast this, the MY 27 dust storm was weaker as whole (Figure~\ref{dustmaps}), and the global dust content was smaller before the event began (it happened later in the year during the second peak of the annual dust cycle).
    
    To conclude, the peak altitude may have increased 10-20 km during the MY 29 dust storm, but the increase is comparable to the exceptionally high variability observed at that time.

    \subsection{Mars Year 32}
    \label{sec:MY32}
    
    Figure~\ref{MY32} summarizes observations from the MY 32 dust season between L$_s$ 215$^{\circ}$-230$^{\circ}$. The observing conditions are very similar to those in MY 29 (Figure~\ref{MY29}). In both cases, the dust level is already elevated when the observing period begins, and the dust optical depths increase by a factor of $\sim$1.5 over $\sim$10$^{\circ}$ of L$_s$. In MY 32, however, the MARSIS observations are now in the northern hemisphere instead of the southern hemisphere, and they cover a higher SZA range between 75$^{\circ}$-80$^{\circ}$.
    
    The peak altitude, shown in Figure~\ref{MY32}c, increases 10-15 km between L$_s$ 221$^{\circ}$-222$^{\circ}$ and is somewhat elevated through L$_s$ 225$^{\circ}$. However, similar to MY 29, is is difficult to conclusively determine if this trend is statistically significant because the peak altitude exhibits substantial variability. The variability metric is $\sigma_{h_{max}}$ = 5 km at L$_s < 221^{\circ}$ and $\sigma_{h_{max}}$ = 7 km at L$_s > 223^{\circ}$. The relative pressure, shown in Figure~\ref{MY32}d, also exhibits substantial variability; it changes by as much as a factor of eight. 
    
    Although the variability is high in MY 32, it is not as high as in MY 29 ($\sigma_{h_{max}}$ $\simeq$ 7 km vs. $\sigma_{h_{max}}$ = 11 km), which might be a consequence of the different latitudes of the observations. Both are from the southern spring season, but the MY 29 data are from the southern hemisphere, and the MY 32 data are from the northern hemisphere.
    
    Another consideration is that the solar EUV irradiance was a factor of two smaller during the MY 29 than during the MY 32 period (2.4 mW/m$^2$ vs. 1.2 mW/m$^2$, Figure~\ref{overview}d). The higher variability observed in MY 29, then, might be explained by increased atmospheric gravity wave activity. Gravity wave activity is anticorrelated with the thermospheric temperature \citep{england2017,terada2017,harada2018}. Given that the solar EUV heating rate was significantly smaller in MY 29, thermospheric temperatures were likely lower \citep{bougher2015,thiemann2018a}, and strong gravity wave perturbations were likely present. These perturbations in the thermospheric pressure would drive significant variations in the ionospheric peak altitude (Equation~\ref{eq5}), making the peak altitudes more variable in MY 29 than in MY 32.
    
    If this hypothesis is correct, we might also expect higher variability in MY 27 when the EUV irradiance was only 1.3 mW/m$^2$. The peak altitude, however, is less variable in MY 27 ($\sigma_{h_{max}}$ = 4 km) than in both MY 29 ($\sigma_{h_{max}}$ = 11 km) and MY 32 ($\sigma_{h_{max}}$ $\simeq$ 7 km).   
    This once again points to differences in the magnitudes of the dust storms.  During the weak MY 27 dust storm, we observe a stable peak altitude that clearly increases in altitude, but during the relatively stronger MY 29 and MY 32 dust storms, we observe highly variable peak altitudes and less significant increases. This comparison suggests dust storms increase upper atmospheric variability, and that the magnitude of the variability is proportional to the strength of the dust storm. 
    
    In summary, the peak altitude may have increased 10-15 km during the MY 32 dust storm, but, like MY 29, the increasing trend is comparable to the orbit-to-orbit variability.

    \subsection{Mars Year 33}
    \label{sec:MY33}
    
    Figure~\ref{MY33} shows a small subset of MARSIS observations from MY 33 between L$_s$ 304$^{\circ}$-310$^{\circ}$. The MARSIS observations, shown in Figure~\ref{MY33}a, are from SZAs 70$^{\circ}$-80$^{\circ}$ and southern latitudes between -20$^{\circ}$ and -40$^{\circ}$. These observations are specifically targeted in our study because \citet{withers2018}, using observations from the MAVEN Radio Occultation Science Experiment (ROSE), reported that the peak altitude increased $\sim$10 km between L$_s$ 305$^{\circ}$-307$^{\circ}$ during a small, localized dust storm near the south pole. The ROSE observations were obtained at similar SZAs ($\sim$75$^{\circ}$), but at higher latitudes (53$^{\circ}$N, Figure~\ref{MY33}a). 
    
    Figures~\ref{MY33}a-\ref{MY33}b show that, although the dust optical depths are slightly elevated compared to non-dust season they were constant throughout this period (Figures~\ref{overview} and \ref{dustmaps}). Nonetheless, the peak altitude shown in Figure~\ref{MY33}c, increases from $\sim$160 km to 175 km between L$_s$ 304$^{\circ}$-306$^{\circ}$, then falls sharply to 140 km by L$_s$ 309$^{\circ}$. The L$_s$ in which the peak altitude rises coincides with the L$_s$ range reported by \citet{withers2018} for the MAVEN ROSE observations, as marked by the red line in Figure~\ref{MY33}c. 
    
    Although our results are consistent with \citet{withers2018}, it is interesting that the dust optical depths used in our work do not show any dust increase during this time. Figure 4b in \citet{withers2018} clearly shows that a small, localized dust storm near the south pole was observed by the Mars Climate Sounder at L$_s$ 305$^{\circ}$. This inconsistency can likely be explained by considering two major differences in the dust data used. First, \citet{withers2018} used MCS dust opacities at the specific pressure level of 50 Pa, while we use MCS estimated column dust optical depths, derived by integrating the MCS dust opacity over all available levels, including over the extrapolated part of the dust opacity profile down to the ground, assuming the dust is well mixed \citep{montabone2015}. Second, recent analysis of MY 34 MCS column dust optical depths have highlighted large differences between dayside and nightside values during dust storms, with dayside values generally being larger \citep{montabone2019}. While these differences are still a topic of ongoing research, the MY 33 reconstructed maps of column dust optical depth used in our work were constructed primarily from nightside MCS observations \citep{montabone2015}. Hence, it is possible that the optical depths used in our work are low-biased such that the small, localized dust storm in MY 33 is missing from the maps.
    
    For completeness, Figure~\ref{MY33}d shows the MCS dust extinction map at L$_s$ 275$^{\circ}$-330$^{\circ}$. We produced the map by bin-averaging the MCS dust extinction data from the 50 Pa level \citep{mccleese2007} over 3$^{\circ}$ in latitude and 1$^{\circ}$ in L$_s$. The map is nearly equivalent to Figure 4b in \citet{withers2018} and confirms the small, localized dust storm near the south pole that started at L$_s$ $\sim$ 305$^{\circ}$. 
    
    
    It is also interesting to consider that the MAVEN ROSE observations are from northern latitudes near $+50^{\circ}$ while the MARSIS observations are from southern latitudes near $-30^{\circ}$ (Fig.~\ref{MY33}a). Despite the large separation between them, both instruments observe the peak altitude rise at nearly the same time. This implies that the dust storm affected the upper atmosphere over large spatial distances on short timescales. Furthermore, compared to MARSIS, the MAVEN ROSE peak altitude descends more slowly after being elevated by the dust storm (see Figure 4 in \citep{withers2018}). This implies that the effects of the dust storm lasted longer in the northern hemisphere than in the southern hemisphere, despite the dust storm being localized near the south pole. 
    
    To summarize, the peak altitude increased $\sim$10 km during the MY 33 local dust storm.
    
    \subsection{Mars Year 34}
    \label{subsec:MY34}
    
    Figure~\ref{MY34} summarizes the MARSIS observations from the MY 34 global dust storm between L$_s$ 180$^{\circ}$-195$^{\circ}$. The observations are somewhat unfavorable because the MEX periapsis segment covered high SZAs and was rapidly evolving towards the nightside. Consequently, the observational period covers from before the onset of the dust storm (L$_s \simeq 185^{\circ}$) to after it became a global event (L$_s \simeq 190^{\circ}$) \citep{montabone2019}. It does not cover the entirety of the storm as dust levels remained highly elevated through $\sim$L$_s$ 240$^{\circ}$ (Figure~\ref{dustmaps}). Given the constraints of the observations, we slightly modify our analysis for this event.
    
    The MARSIS observations, shown in Figure~\ref{MY33}a, are from SZAs 60$^{\circ}$-85$^{\circ}$ and northern polar latitudes between +59$^{\circ}$ and +86$^{\circ}$. The global average dust optical depth, shown in Figure~\ref{MY34}b, begins to rapidly increase at L$_s$ $\sim$185$^{\circ}$, from $\sim$0.15 before dust storm onset, to 0.6 by L$_s$ 197$^{\circ}$. The local average dust optical depth, however, is relatively constant, indicating that the lower atmospheric dust content at northern polar latitudes, where the MARSIS observations are from, did not significantly increase. In Figure~\ref{MY33}b we also show a third type of averaged optical depth, which we derive by averaging the dust data from a confined latitude range between +0$^{\circ}$-45$^{\circ}$. This ``northern hemisphere'' average optical depth is a better indicator of dust at locations near the peak altitude observations. Figure~\ref{MY33}b shows that the northern hemisphere dust optical depth increases more rapidly than the local or global dust optical depths, a consequence of the dust storm having originated in the northern hemisphere near +30$^{\circ}$ \citep{sanchez2019}. 
    
    Figure~\ref{MY34}c highlights the limitations of the SZA and latitude coverage during this period. The beginning of the period contains measurements from the southernmost latitudes ($<$ 65$^{\circ}$) that are closest to where the dust storm originated. Also, after L$_s$ 187$^{\circ}$, the SZA range is skewed towards higher and higher values with each passing orbit. Since the peak altitude is strongly dependent on SZA near the terminator (Eq~\ref{eq1}), we must be careful when analyzing the peak altitudes from this period. Therefore, we calculate orbit-averaged peak altitudes from two different SZA ranges: 60$^{\circ}$-85$^{\circ}$ and 60$^{\circ}$-70$^{\circ}$.
    
    The peak altitude at SZAs between 65$^{\circ}$-85$^{\circ}$, shown in Figure~\ref{MY34}d, rises 10-15 km between L$s$ 185$^{\circ}$-188$^{\circ}$. The increase occurs at the onset of the dust storm when the northern hemisphere dust optical depth increases from 0.15 to 0.35. Following its increase, the peak altitude becomes highly variable. Between L$_s$ 188$^{\circ}$-190$^{\circ}$, the peak altitude decreases from 153 ($\pm$ 14) km to 125 ($\pm$ 6) km, then increases to 148 ($\pm$ 6) km the next orbit. Using Equation~\ref{eq5} with $h_{max_i} = 137$ km, these peak altitude changes correspond to relative pressures of 4.0 ($\pm$ 4.0), 0.4 ($\pm$ 0.2), and 3.0 ($\pm$ 1.0), respectively. The variability metric also increases from  $\sigma_{h_{max}}$ = 5 km (L$_s < 185^{\circ}$) to $\sigma_{h_{max}}$ = 11 km (L$_s > 189^{\circ}$), but this increase is not necessarily meaningful because the metric is being compared across two periods that have vastly different SZA coverage (Figure~\ref{MY34}c). 
    
    The peak altitude at SZAs between 60$^{\circ}$-70$^{\circ}$ is shown in Figure~\ref{MY34}e. In this case, we require five or more peak altitude observations to calculate an orbit-average (as opposed to 10 or more in Figure~\ref{MY34}d). Again, the peak altitude rises 10-15 km between 185$^{\circ}$-188$^{\circ}$, concurrent with the increase in the northern hemisphere dust optical depth. The two different SZA ranges (Figures~\ref{MY34}c-d) provide ample evidence that the peak altitude increased at the onset of the MY 34 global dust storm between L$_s$ 185$^{\circ}$-188$^{\circ}$. 
    
    MAVEN ROSE also observed peak altitudes during the MY 34 global dust storm, but at a later time period between L$_s$ 195$^{\circ}$-270$^{\circ}$ \citep{felici2019}. The observed peak altitudes in both the northern (+50$^{\circ}$) and southern (-20$^{\circ}$) hemispheres were somewhat elevated relative to their expected values. However, in the northern hemisphere, ROSE did not observe the peak altitude abruptly increase as one would expect during a dust storm. \citet{felici2019} suggested that the peak altitude in the northern hemisphere might have already been elevated before the ROSE observations began. Our results our consistent with this scenario since we observe the peak rise between L$_s$ 185$^{\circ}$-188$^{\circ}$, before the first ROSE observation at L$_s$ 195$^{\circ}$.

    In summary, the peak altitude increased $\sim$10-15 km during the MY 34 global dust storm, and its variability increased after shortly after it was elevated.

    \section{Discussion and Conclusions}
    \label{sec:conclusions}

    Table 1 provides a description of the MARSIS observing conditions, the dust storm, and the response of the ionospheric peak altitude for the six case studies that were analyzed in Section~\ref{sec:dustseasons}. Elevated peak altitudes were observed during each of the case studies, although in MY 29 and 32 the increases were less significant because the peak altitudes were highly variable. 
    
    The local dust storm in MY 33, the regional dust storm in MY 27, and the global dust storms in MYs 28 and 34 provided the clearest examples of elevated peak altitudes during dust storms. During the MY 27 regional dust storm, the peak altitude sharply increased by 10-15 km over $\sim5^{\circ}$ of L$_s$, and then slowly decreased back to pre-storm altitudes over $\sim15^{\circ}$ of L$_s$. This rapid rise and slow descent is consistent with previously reported observations of the thermosphere and ionosphere during dust storms \citep{keating1998,lillis2010b,england2012,withers2013a,zurek2017}. During the MY 28 global dust storm, the peak altitude also increased by 10-15 km, but the response was more gradual, occurring over $\sim10^{\circ}$ of L$_s$. The elevated peak altitudes during these storms were caused by thermospheric pressure levels rising in response to solar heating of dust, and the subsequent expansion of the lower atmosphere \citep{hantsch1990,haberle1993,bougher1997,keating1998,bell2007,heavens2011,withers2013a,wolkenberg2018,fang2019a,qin2019}.
   
    During MY 33, the peak altitude increased by 10-15 km at L$_s$ 305$^{\circ}$ and at southern latitudes between -20$^{\circ}$ and -40$^{\circ}$. \citet{withers2018} reported MAVEN ROSE observations showing a similar increase in the peak altitude at the same time, but at northern latitudes near +50$^{\circ}$. They attributed the elevated peak altitude to a small, localized dust storm near the south pole. Since the MARSIS and MAVEN observations were separated by 80$^{\circ}$ in latitude, these concurrent observations suggest that this small dust storm spread quickly and affected the upper atmosphere across large distances. 
    
    Another interesting aspect of these two observations is that, after being elevated by the dust storm, the peak altitude observed by MARSIS descended more rapidly than the peak altitude observed by MAVEN ROSE. Hence, even though the dust storm was localized near the south pole, its effect on the upper atmosphere lasted longer in the northern hemisphere than in the southern hemisphere. This surprising result points to the important role of interhemispheric circulation, which allows localized dust storms to affect the upper atmosphere across the planet \citep{bell2007,withers2013a}. As shown in the model simulations \citep{bell2007,medvedev2013}, meridional circulation -- which transfers energy from the summer hemisphere to the winter hemisphere -- is enhanced by increased dust levels. The MARSIS and MAVEN ROSE observations support this result by showing that a dust storm located near the south pole during southern winter can significantly affect the upper atmosphere in both hemispheres. 

    Finally, from the six case studies we conclude that dust storms significantly enhance upper atmospheric variability. The clearest example being the MY 28 global dust storm, during which the peak altitude variability metric more than doubled. Substantial variability was also observed throughout the entire MY 29 and MY 32 periods, which covered the peak of the annual dust cycle near L$_s$ 230$^{\circ}$. Less variability was observed throughout MY 27 near L$_s$ 315$^{\circ}$, when dust levels were significantly lower. During the MY 34 global dust storm, the peak altitude varied by more than 20 km immediately after rising in altitude. Observations of increased upper atmospheric variability during dust storms is not unprecedented. The Mars Global Surveyor's accelerometer observed a more than 100\% increase in the orbit-to-orbit variability of thermospheric densities during the 1997 Noachis regional dust storm \citep{keating1998,bougher1999c}.
    
    The increased variability strongly suggests that dust storms enhance dynamical processes that couple the  lower atmosphere to the upper atmosphere, such as upward propagating gravity waves or atmospheric tides \citep{bougher1999c,medvedev2013,england2017}. Our results point to the importance of including these processes in global atmospheric models, which can successfully reproduce the neutral density and peak altitude increases observed during dust storms \citep{bougher1999c,bell2007,galindo2010,medvedev2013}, but cannot currently reproduce the observed orbit-to-orbit variability. Improvements to these global models will improve our understanding of how dust storms affect the dynamical processes that link the lower and upper atmospheres, and drive substantial thermospheric perturbations.

\acknowledgments
We thank Joe Groene and Chris Piker for helping us acquire MARSIS data using the das2py software package (\url{https://das2.org/das2py/}). We also thank an anonymous reviewer whose suggestions greatly improved the structure and content of the paper. Z. Girazian gratefully acknowledges Paul Withers, Marianna Felici, and Shane Stone for their valuable discussions. This research was supported by NASA through Contract No. 1560641 with the Jet Propulsion Laboratory. F. N\v{e}mec was supported by MSMT grant LTAUSA17070. The MARSIS, EUV, and dust data compiled for this study can be downloaded at \url{https://www.researchgate.net/publication/335104578_MARSIS_EUV_DUST_DATA} and are publicly available. The complete dust maps can be downloaded at \url{www-mars.lmd.jussieu.fr/mars/dust_climatology/}. The MCS data used in Figure~\ref{MY33}d is publicly available and can be downloaded at \url{https://atmos.nmsu.edu/data_and_services/atmospheres_data/MARS/aerosols.html}.  



\bibliography{References_masterPW.bib}

\begin{thebibliography}{}

\bibitem [\protect \citeauthoryear {%
{Bell}%
, {Bougher}%
\BCBL {}\ \BBA {} {Murphy}%
}{%
{Bell}%
\ \protect \BOthers {.}}{%
{\protect \APACyear {2007}}%
}]{%
bell2007}
\APACinsertmetastar {%
bell2007}%
\begin{APACrefauthors}%
{Bell}, J\BPBI M.%
, {Bougher}, S\BPBI W.%
\BCBL {}\ \BBA {} {Murphy}, J\BPBI R.%
\end{APACrefauthors}%
\unskip\
\newblock
\APACrefYearMonthDay{2007}{}{}.
\newblock
{\BBOQ}\APACrefatitle {{Vertical dust mixing and the interannual variations in
  the {M}ars thermosphere}} {{Vertical dust mixing and the interannual
  variations in the {M}ars thermosphere}}.{\BBCQ}
\newblock
\APACjournalVolNumPages{J. Geophys. Res.}{112}{}{E12002}.
\newblock
\begin{APACrefDOI} \doi{10.1029/2006JE002856} \end{APACrefDOI}
\PrintBackRefs{\CurrentBib}

\bibitem [\protect \citeauthoryear {%
S.~{Bougher}%
\ \protect \BOthers {.}}{%
S.~{Bougher}%
\ \protect \BOthers {.}}{%
{\protect \APACyear {1999}}%
}]{%
bougher1999c}
\APACinsertmetastar {%
bougher1999c}%
\begin{APACrefauthors}%
{Bougher}, S.%
, {Keating}, G.%
, {Zurek}, R.%
, {Murphy}, J.%
, {Haberle}, R.%
, {Hollingsworth}, J.%
\BCBL {}\ \BBA {} {Clancy}, R\BPBI T.%
\end{APACrefauthors}%
\unskip\
\newblock
\APACrefYearMonthDay{1999}{}{}.
\newblock
{\BBOQ}\APACrefatitle {{Mars {G}lobal {S}urveyor aerobraking : atmospheric
  trends and model interpretation}} {{Mars {G}lobal {S}urveyor aerobraking :
  atmospheric trends and model interpretation}}.{\BBCQ}
\newblock
\APACjournalVolNumPages{Adv. Space Res.}{23}{11}{1887-1897}.
\newblock
\begin{APACrefDOI} \doi{10.1016/S0273-1177(99)00272-0} \end{APACrefDOI}
\PrintBackRefs{\CurrentBib}

\bibitem [\protect \citeauthoryear {%
S\BPBI W.~{Bougher}%
, {Cravens}%
, {Grebowsky}%
\BCBL {}\ \BBA {} {Luhmann}%
}{%
S\BPBI W.~{Bougher}%
, {Cravens}%
\BCBL {}\ \protect \BOthers {.}}{%
{\protect \APACyear {2015}}%
}]{%
bougher2015a}
\APACinsertmetastar {%
bougher2015a}%
\begin{APACrefauthors}%
{Bougher}, S\BPBI W.%
, {Cravens}, T\BPBI E.%
, {Grebowsky}, J.%
\BCBL {}\ \BBA {} {Luhmann}, J.%
\end{APACrefauthors}%
\unskip\
\newblock
\APACrefYearMonthDay{2015}{}{}.
\newblock
{\BBOQ}\APACrefatitle {{The {A}eronomy of {M}ars: {C}haracterization by {MAVEN}
  of the {U}pper {A}tmosphere {R}eservoir {T}hat {R}egulates {V}olatile
  {E}scape}} {{The {A}eronomy of {M}ars: {C}haracterization by {MAVEN} of the
  {U}pper {A}tmosphere {R}eservoir {T}hat {R}egulates {V}olatile
  {E}scape}}.{\BBCQ}
\newblock
\APACjournalVolNumPages{Space. Sci. Rev.}{195}{}{423-456}.
\newblock
\begin{APACrefDOI} \doi{10.1007/s11214-014-0053-7} \end{APACrefDOI}
\PrintBackRefs{\CurrentBib}

\bibitem [\protect \citeauthoryear {%
S\BPBI W.~{Bougher}%
, {Engel}%
, {Hinson}%
\BCBL {}\ \BBA {} {Forbes}%
}{%
S\BPBI W.~{Bougher}%
\ \protect \BOthers {.}}{%
{\protect \APACyear {2001}}%
}]{%
bougher2001}
\APACinsertmetastar {%
bougher2001}%
\begin{APACrefauthors}%
{Bougher}, S\BPBI W.%
, {Engel}, S.%
, {Hinson}, D\BPBI P.%
\BCBL {}\ \BBA {} {Forbes}, J\BPBI M.%
\end{APACrefauthors}%
\unskip\
\newblock
\APACrefYearMonthDay{2001}{}{}.
\newblock
{\BBOQ}\APACrefatitle {Mars {G}lobal {S}urveyor {R}adio {S}cience electron
  density profiles: {N}eutral atmosphere implications} {Mars {G}lobal
  {S}urveyor {R}adio {S}cience electron density profiles: {N}eutral atmosphere
  implications}.{\BBCQ}
\newblock
\APACjournalVolNumPages{Geophys. Res. Lett.}{28}{}{3091-3094}.
\newblock
\begin{APACrefDOI} \doi{10.1029/2001GL012884} \end{APACrefDOI}
\PrintBackRefs{\CurrentBib}

\bibitem [\protect \citeauthoryear {%
S\BPBI W.~{Bougher}%
, {Engel}%
, {Hinson}%
\BCBL {}\ \BBA {} {Murphy}%
}{%
S\BPBI W.~{Bougher}%
\ \protect \BOthers {.}}{%
{\protect \APACyear {2004}}%
}]{%
bougher2004}
\APACinsertmetastar {%
bougher2004}%
\begin{APACrefauthors}%
{Bougher}, S\BPBI W.%
, {Engel}, S.%
, {Hinson}, D\BPBI P.%
\BCBL {}\ \BBA {} {Murphy}, J\BPBI R.%
\end{APACrefauthors}%
\unskip\
\newblock
\APACrefYearMonthDay{2004}{}{}.
\newblock
{\BBOQ}\APACrefatitle {{MGS} {R}adio {S}cience electron density profiles:
  {I}nterannual variability and implications for the {M}artian neutral
  atmosphere} {{MGS} {R}adio {S}cience electron density profiles: {I}nterannual
  variability and implications for the {M}artian neutral atmosphere}.{\BBCQ}
\newblock
\APACjournalVolNumPages{J. Geophys. Res.}{109}{}{E03010}.
\newblock
\begin{APACrefDOI} \doi{10.1029/2003JE002154} \end{APACrefDOI}
\PrintBackRefs{\CurrentBib}

\bibitem [\protect \citeauthoryear {%
S\BPBI W.~{Bougher}%
, {Murphy}%
\BCBL {}\ \BBA {} {Haberle}%
}{%
S\BPBI W.~{Bougher}%
\ \protect \BOthers {.}}{%
{\protect \APACyear {1997}}%
}]{%
bougher1997}
\APACinsertmetastar {%
bougher1997}%
\begin{APACrefauthors}%
{Bougher}, S\BPBI W.%
, {Murphy}, J.%
\BCBL {}\ \BBA {} {Haberle}, R\BPBI M.%
\end{APACrefauthors}%
\unskip\
\newblock
\APACrefYearMonthDay{1997}{}{}.
\newblock
{\BBOQ}\APACrefatitle {{Dust storm impacts on the {M}ars upper atmosphere}}
  {{Dust storm impacts on the {M}ars upper atmosphere}}.{\BBCQ}
\newblock
\APACjournalVolNumPages{Adv. Space Res.}{19}{}{1255-1260}.
\newblock
\begin{APACrefDOI} \doi{10.1016/S0273-1177(97)00278-0} \end{APACrefDOI}
\PrintBackRefs{\CurrentBib}

\bibitem [\protect \citeauthoryear {%
S\BPBI W.~{Bougher}%
, {Pawlowski}%
\BCBL {}\ \protect \BOthers {.}}{%
S\BPBI W.~{Bougher}%
, {Pawlowski}%
\BCBL {}\ \protect \BOthers {.}}{%
{\protect \APACyear {2015}}%
}]{%
bougher2015}
\APACinsertmetastar {%
bougher2015}%
\begin{APACrefauthors}%
{Bougher}, S\BPBI W.%
, {Pawlowski}, D.%
, {Bell}, J\BPBI M.%
, {Nelli}, S.%
, {McDunn}, T.%
, {Murphy}, J\BPBI R.%
\BDBL {}{Ridley}, A.%
\end{APACrefauthors}%
\unskip\
\newblock
\APACrefYearMonthDay{2015}{}{}.
\newblock
{\BBOQ}\APACrefatitle {{Mars {G}lobal {I}onosphere-{T}hermosphere {M}odel:
  {S}olar cycle, seasonal, and diurnal variations of the {M}ars upper
  atmosphere}} {{Mars {G}lobal {I}onosphere-{T}hermosphere {M}odel: {S}olar
  cycle, seasonal, and diurnal variations of the {M}ars upper
  atmosphere}}.{\BBCQ}
\newblock
\APACjournalVolNumPages{J. Geophys. Res.}{120}{}{311-342}.
\newblock
\begin{APACrefDOI} \doi{10.1002/2014JE004715} \end{APACrefDOI}
\PrintBackRefs{\CurrentBib}

\bibitem [\protect \citeauthoryear {%
{Breus}%
\ \protect \BOthers {.}}{%
{Breus}%
\ \protect \BOthers {.}}{%
{\protect \APACyear {2004}}%
}]{%
breus2004}
\APACinsertmetastar {%
breus2004}%
\begin{APACrefauthors}%
{Breus}, T\BPBI K.%
, {Krymskii}, A\BPBI M.%
, {Crider}, D\BPBI H.%
, {Ness}, N\BPBI F.%
, {Hinson}, D.%
\BCBL {}\ \BBA {} {Barashyan}, K\BPBI K.%
\end{APACrefauthors}%
\unskip\
\newblock
\APACrefYearMonthDay{2004}{}{}.
\newblock
{\BBOQ}\APACrefatitle {Effect of the solar radiation in the topside
  atmosphere/ionosphere of {M}ars: {M}ars {G}lobal {S}urveyor observations}
  {Effect of the solar radiation in the topside atmosphere/ionosphere of
  {M}ars: {M}ars {G}lobal {S}urveyor observations}.{\BBCQ}
\newblock
\APACjournalVolNumPages{J. Geophys. Res.}{109}{}{A09310}.
\newblock
\begin{APACrefDOI} \doi{10.1029/2004JA010431} \end{APACrefDOI}
\PrintBackRefs{\CurrentBib}

\bibitem [\protect \citeauthoryear {%
{Chaffin}%
, {Deighan}%
, {Schneider}%
\BCBL {}\ \BBA {} {Stewart}%
}{%
{Chaffin}%
\ \protect \BOthers {.}}{%
{\protect \APACyear {2017}}%
}]{%
chaffin2017}
\APACinsertmetastar {%
chaffin2017}%
\begin{APACrefauthors}%
{Chaffin}, M\BPBI S.%
, {Deighan}, J.%
, {Schneider}, N\BPBI M.%
\BCBL {}\ \BBA {} {Stewart}, A\BPBI I\BPBI F.%
\end{APACrefauthors}%
\unskip\
\newblock
\APACrefYearMonthDay{2017}{}{}.
\newblock
{\BBOQ}\APACrefatitle {{Elevated atmospheric escape of atomic hydrogen from
  {M}ars induced by high-altitude water}} {{Elevated atmospheric escape of
  atomic hydrogen from {M}ars induced by high-altitude water}}.{\BBCQ}
\newblock
\APACjournalVolNumPages{Nature Geoscience}{10}{}{174-178}.
\newblock
\begin{APACrefDOI} \doi{10.1038/ngeo2887} \end{APACrefDOI}
\PrintBackRefs{\CurrentBib}

\bibitem [\protect \citeauthoryear {%
Chapman%
}{%
Chapman%
}{%
{\protect \APACyear {1931}}%
}]{%
chapman1931a}
\APACinsertmetastar {%
chapman1931a}%
\begin{APACrefauthors}%
Chapman, S.%
\end{APACrefauthors}%
\unskip\
\newblock
\APACrefYearMonthDay{1931}{}{}.
\newblock
{\BBOQ}\APACrefatitle {The Absorption and Dissociative or Ionizing Effect of
  MonochromaticRadiation in an Atmosphere on a Rotating {E}arth} {The
  absorption and dissociative or ionizing effect of monochromaticradiation in
  an atmosphere on a rotating {E}arth}.{\BBCQ}
\newblock
\APACjournalVolNumPages{Proc. Phys. Soc.}{43}{}{26-45}.
\newblock
\begin{APACrefDOI} \doi{10.1088/0959-5309/43/1/305} \end{APACrefDOI}
\PrintBackRefs{\CurrentBib}

\bibitem [\protect \citeauthoryear {%
{Christensen}%
\ \protect \BOthers {.}}{%
{Christensen}%
\ \protect \BOthers {.}}{%
{\protect \APACyear {2004}}%
}]{%
christensen2004}
\APACinsertmetastar {%
christensen2004}%
\begin{APACrefauthors}%
{Christensen}, P\BPBI R.%
, {Jakosky}, B\BPBI M.%
, {Kieffer}, H\BPBI H.%
, {Malin}, M\BPBI C.%
, {McSween}, H\BPBI Y., Jr.%
, {Nealson}, K.%
\BDBL {}{Ravine}, M.%
\end{APACrefauthors}%
\unskip\
\newblock
\APACrefYearMonthDay{2004}{}{}.
\newblock
{\BBOQ}\APACrefatitle {{The {T}hermal {E}mission {I}maging {S}ystem {(THEMIS)}
  for the {M}ars 2001 {O}dyssey mission}} {{The {T}hermal {E}mission {I}maging
  {S}ystem {(THEMIS)} for the {M}ars 2001 {O}dyssey mission}}.{\BBCQ}
\newblock
\APACjournalVolNumPages{Space. Sci. Rev.}{110}{}{85-130}.
\newblock
\begin{APACrefDOI} \doi{10.1023/B:SPAC.0000021008.16305.94} \end{APACrefDOI}
\PrintBackRefs{\CurrentBib}

\bibitem [\protect \citeauthoryear {%
{Clancy}%
\ \protect \BOthers {.}}{%
{Clancy}%
\ \protect \BOthers {.}}{%
{\protect \APACyear {2010}}%
}]{%
clancy2010}
\APACinsertmetastar {%
clancy2010}%
\begin{APACrefauthors}%
{Clancy}, R\BPBI T.%
, {Wolff}, M\BPBI J.%
, {Whitney}, B\BPBI A.%
, {Cantor}, B\BPBI A.%
, {Smith}, M\BPBI D.%
\BCBL {}\ \BBA {} {McConnochie}, T\BPBI H.%
\end{APACrefauthors}%
\unskip\
\newblock
\APACrefYearMonthDay{2010}{}{}.
\newblock
{\BBOQ}\APACrefatitle {{Extension of atmospheric dust loading to high altitudes
  during the 2001 {M}ars dust storm: {MGS TES} limb observations}} {{Extension
  of atmospheric dust loading to high altitudes during the 2001 {M}ars dust
  storm: {MGS TES} limb observations}}.{\BBCQ}
\newblock
\APACjournalVolNumPages{Icarus}{207}{}{98-109}.
\newblock
\begin{APACrefDOI} \doi{10.1016/j.icarus.2009.10.011} \end{APACrefDOI}
\PrintBackRefs{\CurrentBib}

\bibitem [\protect \citeauthoryear {%
{Daerden}%
\ \protect \BOthers {.}}{%
{Daerden}%
\ \protect \BOthers {.}}{%
{\protect \APACyear {2019}}%
}]{%
daerden2019}
\APACinsertmetastar {%
daerden2019}%
\begin{APACrefauthors}%
{Daerden}, F.%
, {Neary}, L.%
, {Viscardy}, S.%
, {Garc{\'{\i}}a Mu{\~n}oz}, A.%
, {Clancy}, R\BPBI T.%
, {Smith}, M\BPBI D.%
\BDBL {}{Fedorova}, A.%
\end{APACrefauthors}%
\unskip\
\newblock
\APACrefYearMonthDay{2019}{}{}.
\newblock
{\BBOQ}\APACrefatitle {{Mars atmospheric chemistry simulations with the
  {GEM}-{M}ars general circulation model}} {{Mars atmospheric chemistry
  simulations with the {GEM}-{M}ars general circulation model}}.{\BBCQ}
\newblock
\APACjournalVolNumPages{Icarus}{326}{}{197-224}.
\newblock
\begin{APACrefDOI} \doi{10.1016/j.icarus.2019.02.030} \end{APACrefDOI}
\PrintBackRefs{\CurrentBib}

\bibitem [\protect \citeauthoryear {%
{Duru}%
\ \protect \BOthers {.}}{%
{Duru}%
\ \protect \BOthers {.}}{%
{\protect \APACyear {2008}}%
}]{%
duru2008}
\APACinsertmetastar {%
duru2008}%
\begin{APACrefauthors}%
{Duru}, F.%
, {Gurnett}, D\BPBI A.%
, {Morgan}, D\BPBI D.%
, {Modolo}, R.%
, {Nagy}, A\BPBI F.%
\BCBL {}\ \BBA {} {Najib}, D.%
\end{APACrefauthors}%
\unskip\
\newblock
\APACrefYearMonthDay{2008}{}{}.
\newblock
{\BBOQ}\APACrefatitle {Electron densities in the upper ionosphere of {M}ars
  from the excitation of electron plasma oscillations} {Electron densities in
  the upper ionosphere of {M}ars from the excitation of electron plasma
  oscillations}.{\BBCQ}
\newblock
\APACjournalVolNumPages{J. Geophys. Res.}{113}{}{A07302}.
\newblock
\begin{APACrefDOI} \doi{10.1029/2008JA013073} \end{APACrefDOI}
\PrintBackRefs{\CurrentBib}

\bibitem [\protect \citeauthoryear {%
{England}%
\ \BBA {} {Lillis}%
}{%
{England}%
\ \BBA {} {Lillis}%
}{%
{\protect \APACyear {2012}}%
}]{%
england2012}
\APACinsertmetastar {%
england2012}%
\begin{APACrefauthors}%
{England}, S\BPBI L.%
\BCBT {}\ \BBA {} {Lillis}, R\BPBI J.%
\end{APACrefauthors}%
\unskip\
\newblock
\APACrefYearMonthDay{2012}{}{}.
\newblock
{\BBOQ}\APACrefatitle {{On the nature of the variability of the {M}artian
  thermospheric mass density: {R}esults from electron reflectometry with {M}ars
  {G}lobal {S}urveyor}} {{On the nature of the variability of the {M}artian
  thermospheric mass density: {R}esults from electron reflectometry with {M}ars
  {G}lobal {S}urveyor}}.{\BBCQ}
\newblock
\APACjournalVolNumPages{J. Geophys. Res.}{117}{E2}{E02008}.
\newblock
\begin{APACrefDOI} \doi{10.1029/2011JE003998} \end{APACrefDOI}
\PrintBackRefs{\CurrentBib}

\bibitem [\protect \citeauthoryear {%
{England}%
\ \protect \BOthers {.}}{%
{England}%
\ \protect \BOthers {.}}{%
{\protect \APACyear {2017}}%
}]{%
england2017}
\APACinsertmetastar {%
england2017}%
\begin{APACrefauthors}%
{England}, S\BPBI L.%
, {Liu}, G.%
, {Yi{\v{g}}it}, E.%
, {Mahaffy}, P\BPBI R.%
, {Elrod}, M.%
, {Benna}, M.%
\BDBL {}{Jakosky}, B.%
\end{APACrefauthors}%
\unskip\
\newblock
\APACrefYearMonthDay{2017}{}{}.
\newblock
{\BBOQ}\APACrefatitle {{{MAVEN NGIMS} observations of atmospheric gravity waves
  in the {Martian} thermosphere}} {{{MAVEN NGIMS} observations of atmospheric
  gravity waves in the {Martian} thermosphere}}.{\BBCQ}
\newblock
\APACjournalVolNumPages{J. Geophys. Res.}{122}{2}{2310-2335}.
\PrintBackRefs{\CurrentBib}

\bibitem [\protect \citeauthoryear {%
{Fallows}%
, {Withers}%
\BCBL {}\ \BBA {} {Matta}%
}{%
{Fallows}%
\ \protect \BOthers {.}}{%
{\protect \APACyear {2015}}%
}]{%
fallows2015}
\APACinsertmetastar {%
fallows2015}%
\begin{APACrefauthors}%
{Fallows}, K.%
, {Withers}, P.%
\BCBL {}\ \BBA {} {Matta}, M.%
\end{APACrefauthors}%
\unskip\
\newblock
\APACrefYearMonthDay{2015}{}{}.
\newblock
{\BBOQ}\APACrefatitle {{An observational study of the influence of solar zenith
  angle on properties of the {M}1 layer of the {M}ars ionosphere}} {{An
  observational study of the influence of solar zenith angle on properties of
  the {M}1 layer of the {M}ars ionosphere}}.{\BBCQ}
\newblock
\APACjournalVolNumPages{J. Geophys. Res.}{120}{}{1299-1310}.
\newblock
\begin{APACrefDOI} \doi{10.1002/2014JA020750} \end{APACrefDOI}
\PrintBackRefs{\CurrentBib}

\bibitem [\protect \citeauthoryear {%
Fang%
\ \protect \BOthers {.}}{%
Fang%
\ \protect \BOthers {.}}{%
{\protect \APACyear {2019}}%
}]{%
fang2019a}
\APACinsertmetastar {%
fang2019a}%
\begin{APACrefauthors}%
Fang, X.%
, Ma, Y.%
, Lee, Y.%
, Bougher, S.%
, Liu, G.%
, Benna, M.%
\BDBL {}Jakosky, B.%
\end{APACrefauthors}%
\unskip\
\newblock
\APACrefYearMonthDay{2019}{}{}.
\newblock
{\BBOQ}\APACrefatitle {Mars Dust Storm Effects in the Ionosphere and
  Magnetosphere and Implications for Atmospheric Carbon Loss} {Mars dust storm
  effects in the ionosphere and magnetosphere and implications for atmospheric
  carbon loss}.{\BBCQ}
\newblock
\APACjournalVolNumPages{J. Geophys. Res.}{Accepted}{}{}.
\newblock
\begin{APACrefDOI} \doi{10.1029/2019JA026838} \end{APACrefDOI}
\PrintBackRefs{\CurrentBib}

\bibitem [\protect \citeauthoryear {%
{Felici}%
, {Withers}%
\BCBL {}\ \BBA {} {Montabone}%
}{%
{Felici}%
\ \protect \BOthers {.}}{%
{\protect \APACyear {2019, submitted to this issue}}%
}]{%
felici2019}
\APACinsertmetastar {%
felici2019}%
\begin{APACrefauthors}%
{Felici}, M.%
, {Withers}, P.%
\BCBL {}\ \BBA {} {Montabone}, L.%
\end{APACrefauthors}%
\unskip\
\newblock
\APACrefYearMonthDay{2019, submitted to this issue}{}{}.
\newblock
{\BBOQ}\APACrefatitle {{MAVEN ROSE observations of the response of the
  {M}artian ionosphere to dust storms}} {{MAVEN ROSE observations of the
  response of the {M}artian ionosphere to dust storms}}.{\BBCQ}
\newblock
\APACjournalVolNumPages{Submitted to J. Geophys. Res.}{}{}{}.
\PrintBackRefs{\CurrentBib}

\bibitem [\protect \citeauthoryear {%
{Fox}%
\ \BBA {} {Weber}%
}{%
{Fox}%
\ \BBA {} {Weber}%
}{%
{\protect \APACyear {2012}}%
}]{%
fox2012}
\APACinsertmetastar {%
fox2012}%
\begin{APACrefauthors}%
{Fox}, J\BPBI L.%
\BCBT {}\ \BBA {} {Weber}, A\BPBI J.%
\end{APACrefauthors}%
\unskip\
\newblock
\APACrefYearMonthDay{2012}{{\APACmonth{11}}}{}.
\newblock
{\BBOQ}\APACrefatitle {{MGS} electron density profiles: {A}nalysis and modeling
  of peak altitudes} {{MGS} electron density profiles: {A}nalysis and modeling
  of peak altitudes}.{\BBCQ}
\newblock
\APACjournalVolNumPages{Icarus}{221}{}{1002-1019}.
\newblock
\begin{APACrefDOI} \doi{10.1016/j.icarus.2012.10.002} \end{APACrefDOI}
\PrintBackRefs{\CurrentBib}

\bibitem [\protect \citeauthoryear {%
{Girazian}%
\ \BBA {} {Withers}%
}{%
{Girazian}%
\ \BBA {} {Withers}%
}{%
{\protect \APACyear {2013}}%
}]{%
girazian2013}
\APACinsertmetastar {%
girazian2013}%
\begin{APACrefauthors}%
{Girazian}, Z.%
\BCBT {}\ \BBA {} {Withers}, P.%
\end{APACrefauthors}%
\unskip\
\newblock
\APACrefYearMonthDay{2013}{}{}.
\newblock
{\BBOQ}\APACrefatitle {{The dependence of peak electron density in the
  ionosphere of {M}ars on solar irradiance}} {{The dependence of peak electron
  density in the ionosphere of {M}ars on solar irradiance}}.{\BBCQ}
\newblock
\APACjournalVolNumPages{Geophys. Res. Lett.}{40}{}{1960-1964}.
\newblock
\begin{APACrefDOI} \doi{10.1002/grl.50344} \end{APACrefDOI}
\PrintBackRefs{\CurrentBib}

\bibitem [\protect \citeauthoryear {%
{Girazian}%
\ \BBA {} {Withers}%
}{%
{Girazian}%
\ \BBA {} {Withers}%
}{%
{\protect \APACyear {2015}}%
}]{%
girazian2015b}
\APACinsertmetastar {%
girazian2015b}%
\begin{APACrefauthors}%
{Girazian}, Z.%
\BCBT {}\ \BBA {} {Withers}, P.%
\end{APACrefauthors}%
\unskip\
\newblock
\APACrefYearMonthDay{2015}{}{}.
\newblock
{\BBOQ}\APACrefatitle {{An empirical model of the extreme ultraviolet solar
  spectrum as a function of {F}10.7}} {{An empirical model of the extreme
  ultraviolet solar spectrum as a function of {F}10.7}}.{\BBCQ}
\newblock
\APACjournalVolNumPages{J. Geophys. Res.}{120}{}{6779–6794}.
\newblock
\begin{APACrefDOI} \doi{10.1002/2015JA021436} \end{APACrefDOI}
\PrintBackRefs{\CurrentBib}

\bibitem [\protect \citeauthoryear {%
{Gonz{\'a}lez-Galindo}%
, {Bougher}%
, {L{\'o}pez-Valverde}%
, {Forget}%
\BCBL {}\ \BBA {} {Murphy}%
}{%
{Gonz{\'a}lez-Galindo}%
\ \protect \BOthers {.}}{%
{\protect \APACyear {2010}}%
}]{%
galindo2010}
\APACinsertmetastar {%
galindo2010}%
\begin{APACrefauthors}%
{Gonz{\'a}lez-Galindo}, F.%
, {Bougher}, S\BPBI W.%
, {L{\'o}pez-Valverde}, M\BPBI A.%
, {Forget}, F.%
\BCBL {}\ \BBA {} {Murphy}, J.%
\end{APACrefauthors}%
\unskip\
\newblock
\APACrefYearMonthDay{2010}{}{}.
\newblock
{\BBOQ}\APACrefatitle {{Thermal and wind structure of the {M}artian
  thermosphere as given by two {G}eneral {C}irculation {M}odels}} {{Thermal and
  wind structure of the {M}artian thermosphere as given by two {G}eneral
  {C}irculation {M}odels}}.{\BBCQ}
\newblock
\APACjournalVolNumPages{Planet. Space Sci.}{58}{}{1832-1849}.
\newblock
\begin{APACrefDOI} \doi{10.1016/j.pss.2010.08.013} \end{APACrefDOI}
\PrintBackRefs{\CurrentBib}

\bibitem [\protect \citeauthoryear {%
{Gurnett}%
\ \protect \BOthers {.}}{%
{Gurnett}%
\ \protect \BOthers {.}}{%
{\protect \APACyear {2008}}%
}]{%
gurnett2008}
\APACinsertmetastar {%
gurnett2008}%
\begin{APACrefauthors}%
{Gurnett}, D\BPBI A.%
, {Huff}, R\BPBI L.%
, {Morgan}, D\BPBI D.%
, {Persoon}, A\BPBI M.%
, {Averkamp}, T\BPBI F.%
, {Kirchner}, D\BPBI L.%
\BDBL {}{Picardi}, G.%
\end{APACrefauthors}%
\unskip\
\newblock
\APACrefYearMonthDay{2008}{}{}.
\newblock
{\BBOQ}\APACrefatitle {An overview of radar soundings of the martian ionosphere
  from the {M}ars {E}xpress spacecraft} {An overview of radar soundings of the
  martian ionosphere from the {M}ars {E}xpress spacecraft}.{\BBCQ}
\newblock
\APACjournalVolNumPages{Adv. Space Res.}{41}{}{1335-1346}.
\newblock
\begin{APACrefDOI} \doi{10.1016/j.asr.2007.01.062} \end{APACrefDOI}
\PrintBackRefs{\CurrentBib}

\bibitem [\protect \citeauthoryear {%
{Gurnett}%
\ \protect \BOthers {.}}{%
{Gurnett}%
\ \protect \BOthers {.}}{%
{\protect \APACyear {2005}}%
}]{%
gurnett2005}
\APACinsertmetastar {%
gurnett2005}%
\begin{APACrefauthors}%
{Gurnett}, D\BPBI A.%
, {Kirchner}, D\BPBI L.%
, {Huff}, R\BPBI L.%
, {Morgan}, D\BPBI D.%
, {Persoon}, A\BPBI M.%
, {Averkamp}, T\BPBI F.%
\BDBL {}{Picardi}, G.%
\end{APACrefauthors}%
\unskip\
\newblock
\APACrefYearMonthDay{2005}{}{}.
\newblock
{\BBOQ}\APACrefatitle {{Radar Soundings of the Ionosphere of {M}ars}} {{Radar
  Soundings of the Ionosphere of {M}ars}}.{\BBCQ}
\newblock
\APACjournalVolNumPages{Science}{310}{5756}{1929-1933}.
\newblock
\begin{APACrefDOI} \doi{10.1126/science.1121868} \end{APACrefDOI}
\PrintBackRefs{\CurrentBib}

\bibitem [\protect \citeauthoryear {%
Guzewich%
\ \protect \BOthers {.}}{%
Guzewich%
\ \protect \BOthers {.}}{%
{\protect \APACyear {2019}}%
}]{%
guzewich2019}
\APACinsertmetastar {%
guzewich2019}%
\begin{APACrefauthors}%
Guzewich, S\BPBI D.%
, Lemmon, M.%
, Smith, C\BPBI L.%
, Martínez, G.%
, de Vicente-Retortillo, Ã.%
, Newman, C\BPBI E.%
\BDBL {}Zorzano~Mier, M\BHBI P.%
\end{APACrefauthors}%
\unskip\
\newblock
\APACrefYearMonthDay{2019}{}{}.
\newblock
{\BBOQ}\APACrefatitle {Mars {S}cience {L}aboratory Observations of the
  2018/{M}ars {Y}ear 34 Global Dust Storm} {Mars {S}cience {L}aboratory
  observations of the 2018/{M}ars {Y}ear 34 global dust storm}.{\BBCQ}
\newblock
\APACjournalVolNumPages{Geophys. Res. Lett.}{46}{1}{71-79}.
\newblock
\begin{APACrefDOI} \doi{10.1029/2018GL080839} \end{APACrefDOI}
\PrintBackRefs{\CurrentBib}

\bibitem [\protect \citeauthoryear {%
{Haberle}%
\ \protect \BOthers {.}}{%
{Haberle}%
\ \protect \BOthers {.}}{%
{\protect \APACyear {1993}}%
}]{%
haberle1993}
\APACinsertmetastar {%
haberle1993}%
\begin{APACrefauthors}%
{Haberle}, R\BPBI M.%
, {Pollack}, J\BPBI B.%
, {Barnes}, J\BPBI R.%
, {Zurek}, R\BPBI W.%
, {Leovy}, C\BPBI B.%
, {Murphy}, J\BPBI R.%
\BDBL {}{Schaeffer}, J.%
\end{APACrefauthors}%
\unskip\
\newblock
\APACrefYearMonthDay{1993}{}{}.
\newblock
{\BBOQ}\APACrefatitle {{Mars atmospheric dynamics as simulated by the {NASA
  AMES} {G}eneral {C}irculation {M}odel. {I} - {T}he zonal-mean circulation}}
  {{Mars atmospheric dynamics as simulated by the {NASA AMES} {G}eneral
  {C}irculation {M}odel. {I} - {T}he zonal-mean circulation}}.{\BBCQ}
\newblock
\APACjournalVolNumPages{J. Geophys. Res.}{98}{}{3093-3123}.
\newblock
\begin{APACrefDOI} \doi{10.1029/92JE02946} \end{APACrefDOI}
\PrintBackRefs{\CurrentBib}

\bibitem [\protect \citeauthoryear {%
{Hantsch}%
\ \BBA {} {Bauer}%
}{%
{Hantsch}%
\ \BBA {} {Bauer}%
}{%
{\protect \APACyear {1990}}%
}]{%
hantsch1990}
\APACinsertmetastar {%
hantsch1990}%
\begin{APACrefauthors}%
{Hantsch}, M\BPBI H.%
\BCBT {}\ \BBA {} {Bauer}, S\BPBI J.%
\end{APACrefauthors}%
\unskip\
\newblock
\APACrefYearMonthDay{1990}{}{}.
\newblock
{\BBOQ}\APACrefatitle {{Solar control of the {M}ars ionosphere}} {{Solar
  control of the {M}ars ionosphere}}.{\BBCQ}
\newblock
\APACjournalVolNumPages{Planet. Space Sci.}{38}{}{539-542}.
\newblock
\begin{APACrefDOI} \doi{10.1016/0032-0633(90)90146-H} \end{APACrefDOI}
\PrintBackRefs{\CurrentBib}

\bibitem [\protect \citeauthoryear {%
{Harada}%
, {Gurnett}%
, {Kopf}%
, {Halekas}%
\BCBL {}\ \BBA {} {Ruhunusiri}%
}{%
{Harada}%
\ \protect \BOthers {.}}{%
{\protect \APACyear {2018}}%
}]{%
harada2018}
\APACinsertmetastar {%
harada2018}%
\begin{APACrefauthors}%
{Harada}, Y.%
, {Gurnett}, D\BPBI A.%
, {Kopf}, A\BPBI J.%
, {Halekas}, J\BPBI S.%
\BCBL {}\ \BBA {} {Ruhunusiri}, S.%
\end{APACrefauthors}%
\unskip\
\newblock
\APACrefYearMonthDay{2018}{}{}.
\newblock
{\BBOQ}\APACrefatitle {{Ionospheric Irregularities at {M}ars Probed by {MARSIS}
  Topside Sounding}} {{Ionospheric Irregularities at {M}ars Probed by {MARSIS}
  Topside Sounding}}.{\BBCQ}
\newblock
\APACjournalVolNumPages{J. Geophys. Res.}{123}{1}{1018-1030}.
\newblock
\begin{APACrefDOI} \doi{10.1002/2017JA024913} \end{APACrefDOI}
\PrintBackRefs{\CurrentBib}

\bibitem [\protect \citeauthoryear {%
{Heavens}%
\ \protect \BOthers {.}}{%
{Heavens}%
\ \protect \BOthers {.}}{%
{\protect \APACyear {2018}}%
}]{%
heavens2018}
\APACinsertmetastar {%
heavens2018}%
\begin{APACrefauthors}%
{Heavens}, N\BPBI G.%
, {Kleinb{\"o}hl}, A.%
, {Chaffin}, M\BPBI S.%
, {Halekas}, J\BPBI S.%
, {Kass}, D\BPBI M.%
, {Hayne}, P\BPBI O.%
\BDBL {}{Schofield}, J\BPBI T.%
\end{APACrefauthors}%
\unskip\
\newblock
\APACrefYearMonthDay{2018}{}{}.
\newblock
{\BBOQ}\APACrefatitle {{Hydrogen escape from {M}ars enhanced by deep convection
  in dust storms}} {{Hydrogen escape from {M}ars enhanced by deep convection in
  dust storms}}.{\BBCQ}
\newblock
\APACjournalVolNumPages{Nat. Astron.}{2}{}{126-132}.
\newblock
\begin{APACrefDOI} \doi{10.1038/s41550-017-0353-4} \end{APACrefDOI}
\PrintBackRefs{\CurrentBib}

\bibitem [\protect \citeauthoryear {%
{Heavens}%
\ \protect \BOthers {.}}{%
{Heavens}%
\ \protect \BOthers {.}}{%
{\protect \APACyear {2011}}%
}]{%
heavens2011}
\APACinsertmetastar {%
heavens2011}%
\begin{APACrefauthors}%
{Heavens}, N\BPBI G.%
, {McCleese}, D\BPBI J.%
, {Richardson}, M\BPBI I.%
, {Kass}, D\BPBI M.%
, {Kleinb{\"o}hl}, A.%
\BCBL {}\ \BBA {} {Schofield}, J\BPBI T.%
\end{APACrefauthors}%
\unskip\
\newblock
\APACrefYearMonthDay{2011}{}{}.
\newblock
{\BBOQ}\APACrefatitle {{Structure and dynamics of the {M}artian lower and
  middle atmosphere as observed by the {M}ars {C}limate {S}ounder: 2.
  {I}mplications of the thermal structure and aerosol distributions for the
  mean meridional circulation}} {{Structure and dynamics of the {M}artian lower
  and middle atmosphere as observed by the {M}ars {C}limate {S}ounder: 2.
  {I}mplications of the thermal structure and aerosol distributions for the
  mean meridional circulation}}.{\BBCQ}
\newblock
\APACjournalVolNumPages{J. Geophys. Res.}{116}{}{E01010}.
\newblock
\begin{APACrefDOI} \doi{10.1029/2010JE003713} \end{APACrefDOI}
\PrintBackRefs{\CurrentBib}

\bibitem [\protect \citeauthoryear {%
{Keating}%
\ \protect \BOthers {.}}{%
{Keating}%
\ \protect \BOthers {.}}{%
{\protect \APACyear {1998}}%
}]{%
keating1998}
\APACinsertmetastar {%
keating1998}%
\begin{APACrefauthors}%
{Keating}, G\BPBI M.%
, {Bougher}, S\BPBI W.%
, {Zurek}, R\BPBI W.%
, {Tolson}, R\BPBI H.%
, {Cancro}, G\BPBI J.%
, {Noll}, S\BPBI N.%
\BDBL {}{Babicke}, J\BPBI M.%
\end{APACrefauthors}%
\unskip\
\newblock
\APACrefYearMonthDay{1998}{}{}.
\newblock
{\BBOQ}\APACrefatitle {{The Structure of the Upper Atmosphere of Mars: In Situ
  Accelerometer Measurements from Mars Global Surveyor}} {{The Structure of the
  Upper Atmosphere of Mars: In Situ Accelerometer Measurements from Mars Global
  Surveyor}}.{\BBCQ}
\newblock
\APACjournalVolNumPages{Science}{279}{}{1672}.
\newblock
\begin{APACrefDOI} \doi{10.1126/science.279.5357.1672} \end{APACrefDOI}
\PrintBackRefs{\CurrentBib}

\bibitem [\protect \citeauthoryear {%
{Krasnopolsky}%
}{%
{Krasnopolsky}%
}{%
{\protect \APACyear {2019}}%
}]{%
krasnopolsky2019}
\APACinsertmetastar {%
krasnopolsky2019}%
\begin{APACrefauthors}%
{Krasnopolsky}, V\BPBI A.%
\end{APACrefauthors}%
\unskip\
\newblock
\APACrefYearMonthDay{2019}{}{}.
\newblock
{\BBOQ}\APACrefatitle {{Photochemistry of water in the martian thermosphere and
  its effect on hydrogen escape}} {{Photochemistry of water in the martian
  thermosphere and its effect on hydrogen escape}}.{\BBCQ}
\newblock
\APACjournalVolNumPages{Icarus}{321}{}{62-70}.
\newblock
\begin{APACrefDOI} \doi{10.1016/j.icarus.2018.10.033} \end{APACrefDOI}
\PrintBackRefs{\CurrentBib}

\bibitem [\protect \citeauthoryear {%
{Lillis}%
, {Bougher}%
, {Forget}%
, {Smith}%
\BCBL {}\ \BBA {} {Chamberlin}%
}{%
{Lillis}%
\ \protect \BOthers {.}}{%
{\protect \APACyear {2010}}%
}]{%
lillis2010b}
\APACinsertmetastar {%
lillis2010b}%
\begin{APACrefauthors}%
{Lillis}, R\BPBI J.%
, {Bougher}, F., Stephen W.~and{Gonz{\'a}lez-Galindo}%
, {Forget}, F.%
, {Smith}, M\BPBI D.%
\BCBL {}\ \BBA {} {Chamberlin}, P\BPBI C.%
\end{APACrefauthors}%
\unskip\
\newblock
\APACrefYearMonthDay{2010}{}{}.
\newblock
{\BBOQ}\APACrefatitle {{Four {M}artian years of nightside upper thermospheric
  mass densities derived from electron reflectometry: {M}ethod extension and
  comparison with {GCM} simulations}} {{Four {M}artian years of nightside upper
  thermospheric mass densities derived from electron reflectometry: {M}ethod
  extension and comparison with {GCM} simulations}}.{\BBCQ}
\newblock
\APACjournalVolNumPages{J. Geophys. Res.}{115}{E7}{E07014}.
\newblock
\begin{APACrefDOI} \doi{10.1029/2009JE003529} \end{APACrefDOI}
\PrintBackRefs{\CurrentBib}

\bibitem [\protect \citeauthoryear {%
{Liu}%
\ \protect \BOthers {.}}{%
{Liu}%
\ \protect \BOthers {.}}{%
{\protect \APACyear {2018}}%
}]{%
liu2018}
\APACinsertmetastar {%
liu2018}%
\begin{APACrefauthors}%
{Liu}, G.%
, {England}, S\BPBI L.%
, {Lillis}, R\BPBI J.%
, {Withers}, P.%
, {Mahaffy}, P\BPBI R.%
, {Rowland}, D\BPBI E.%
\BDBL {}{Jakosky}, B.%
\end{APACrefauthors}%
\unskip\
\newblock
\APACrefYearMonthDay{2018}{}{}.
\newblock
{\BBOQ}\APACrefatitle {{Thermospheric {E}xpansion {A}ssociated {W}ith {D}ust
  {I}ncrease in the {L}ower {A}tmosphere on {M}ars {O}bserved by
  {MAVEN/NGIMS}}} {{Thermospheric {E}xpansion {A}ssociated {W}ith {D}ust
  {I}ncrease in the {L}ower {A}tmosphere on {M}ars {O}bserved by
  {MAVEN/NGIMS}}}.{\BBCQ}
\newblock
\APACjournalVolNumPages{Geophys. Res. Lett.}{45}{}{2901-2910}.
\newblock
\begin{APACrefDOI} \doi{10.1002/2018GL077525} \end{APACrefDOI}
\PrintBackRefs{\CurrentBib}

\bibitem [\protect \citeauthoryear {%
{Mahaffy}%
\ \protect \BOthers {.}}{%
{Mahaffy}%
\ \protect \BOthers {.}}{%
{\protect \APACyear {2015}}%
}]{%
mahaffy2015a}
\APACinsertmetastar {%
mahaffy2015a}%
\begin{APACrefauthors}%
{Mahaffy}, P\BPBI R.%
, {Benna}, M.%
, {Elrod}, M.%
, {Yelle}, R\BPBI V.%
, {Bougher}, S\BPBI W.%
, {Stone}, S\BPBI W.%
\BCBL {}\ \BBA {} {Jakosky}, B\BPBI M.%
\end{APACrefauthors}%
\unskip\
\newblock
\APACrefYearMonthDay{2015}{}{}.
\newblock
{\BBOQ}\APACrefatitle {{Structure and composition of the neutral upper
  atmosphere of Mars from the {MAVEN} {NGIMS} investigation}} {{Structure and
  composition of the neutral upper atmosphere of Mars from the {MAVEN} {NGIMS}
  investigation}}.{\BBCQ}
\newblock
\APACjournalVolNumPages{Geophys. Res. Lett.}{42}{}{8951-8957}.
\newblock
\begin{APACrefDOI} \doi{10.1002/2015GL065329} \end{APACrefDOI}
\PrintBackRefs{\CurrentBib}

\bibitem [\protect \citeauthoryear {%
{McCleese}%
\ \protect \BOthers {.}}{%
{McCleese}%
\ \protect \BOthers {.}}{%
{\protect \APACyear {2007}}%
}]{%
mccleese2007}
\APACinsertmetastar {%
mccleese2007}%
\begin{APACrefauthors}%
{McCleese}, D\BPBI J.%
, {Schofield}, J\BPBI T.%
, {Taylor}, F\BPBI W.%
, {Calcutt}, S\BPBI B.%
, {Foote}, M\BPBI C.%
, {Kass}, D\BPBI M.%
\BDBL {}{Zurek}, R\BPBI W.%
\end{APACrefauthors}%
\unskip\
\newblock
\APACrefYearMonthDay{2007}{}{}.
\newblock
{\BBOQ}\APACrefatitle {{Mars Climate Sounder: An investigation of thermal and
  water vapor structure, dust and condensate distributions in the atmosphere,
  and energy balance of the polar regions}} {{Mars Climate Sounder: An
  investigation of thermal and water vapor structure, dust and condensate
  distributions in the atmosphere, and energy balance of the polar
  regions}}.{\BBCQ}
\newblock
\APACjournalVolNumPages{J. Geophys. Res.}{112}{}{E05S06}.
\newblock
\begin{APACrefDOI} \doi{10.1029/2006JE002790} \end{APACrefDOI}
\PrintBackRefs{\CurrentBib}

\bibitem [\protect \citeauthoryear {%
{McElroy}%
, {Kong}%
\BCBL {}\ \BBA {} {Yung}%
}{%
{McElroy}%
\ \protect \BOthers {.}}{%
{\protect \APACyear {1977}}%
}]{%
mcelroy1977}
\APACinsertmetastar {%
mcelroy1977}%
\begin{APACrefauthors}%
{McElroy}, M\BPBI B.%
, {Kong}, T\BPBI Y.%
\BCBL {}\ \BBA {} {Yung}, Y\BPBI L.%
\end{APACrefauthors}%
\unskip\
\newblock
\APACrefYearMonthDay{1977}{}{}.
\newblock
{\BBOQ}\APACrefatitle {{Photochemistry and evolution of {M}ars' atmosphere: {A}
  {V}iking perspective}} {{Photochemistry and evolution of {M}ars' atmosphere:
  {A} {V}iking perspective}}.{\BBCQ}
\newblock
\APACjournalVolNumPages{J. Geophys. Res.}{82}{B28}{4379-4388}.
\newblock
\begin{APACrefDOI} \doi{10.1029/JS082i028p04379} \end{APACrefDOI}
\PrintBackRefs{\CurrentBib}

\bibitem [\protect \citeauthoryear {%
{Medvedev}%
, {Yi{\v{g}}it}%
, {Kuroda}%
\BCBL {}\ \BBA {} {Hartogh}%
}{%
{Medvedev}%
\ \protect \BOthers {.}}{%
{\protect \APACyear {2013}}%
}]{%
medvedev2013}
\APACinsertmetastar {%
medvedev2013}%
\begin{APACrefauthors}%
{Medvedev}, A\BPBI S.%
, {Yi{\v{g}}it}, E.%
, {Kuroda}, T.%
\BCBL {}\ \BBA {} {Hartogh}, P.%
\end{APACrefauthors}%
\unskip\
\newblock
\APACrefYearMonthDay{2013}{}{}.
\newblock
{\BBOQ}\APACrefatitle {{General circulation modeling of the Martian upper
  atmosphere during global dust storms}} {{General circulation modeling of the
  Martian upper atmosphere during global dust storms}}.{\BBCQ}
\newblock
\APACjournalVolNumPages{J. Geophys. Res.}{118}{10}{2234-2246}.
\newblock
\begin{APACrefDOI} \doi{10.1002/2013JE004429} \end{APACrefDOI}
\PrintBackRefs{\CurrentBib}

\bibitem [\protect \citeauthoryear {%
{Mendillo}%
\ \protect \BOthers {.}}{%
{Mendillo}%
\ \protect \BOthers {.}}{%
{\protect \APACyear {2017}}%
}]{%
mendillo2017}
\APACinsertmetastar {%
mendillo2017}%
\begin{APACrefauthors}%
{Mendillo}, M.%
, {Narvaez}, C.%
, {Vogt}, M\BPBI F.%
, {Mayyasi}, M.%
, {Forbes}, J.%
, {Galand}, M.%
\BDBL {}{Andersson}, L.%
\end{APACrefauthors}%
\unskip\
\newblock
\APACrefYearMonthDay{2017}{}{}.
\newblock
{\BBOQ}\APACrefatitle {{Sources of Ionospheric Variability at {M}ars}}
  {{Sources of Ionospheric Variability at {M}ars}}.{\BBCQ}
\newblock
\APACjournalVolNumPages{J. Geophys. Res.}{122}{9}{9670-9684}.
\newblock
\begin{APACrefDOI} \doi{10.1002/2017JA024366} \end{APACrefDOI}
\PrintBackRefs{\CurrentBib}

\bibitem [\protect \citeauthoryear {%
{Montabone}%
\ \protect \BOthers {.}}{%
{Montabone}%
\ \protect \BOthers {.}}{%
{\protect \APACyear {2015}}%
}]{%
montabone2015}
\APACinsertmetastar {%
montabone2015}%
\begin{APACrefauthors}%
{Montabone}, L.%
, {Forget}, F.%
, {Millour}, E.%
, {Wilson}, R\BPBI J.%
, {Lewis}, S\BPBI R.%
, {Cantor}, B.%
\BDBL {}{Wolff}, M\BPBI J.%
\end{APACrefauthors}%
\unskip\
\newblock
\APACrefYearMonthDay{2015}{}{}.
\newblock
{\BBOQ}\APACrefatitle {{Eight-year climatology of dust optical depth on
  {M}ars}} {{Eight-year climatology of dust optical depth on {M}ars}}.{\BBCQ}
\newblock
\APACjournalVolNumPages{Icarus}{251}{}{65-95}.
\newblock
\begin{APACrefDOI} \doi{10.1016/j.icarus.2014.12.034} \end{APACrefDOI}
\PrintBackRefs{\CurrentBib}

\bibitem [\protect \citeauthoryear {%
{Montabone}%
\ \protect \BOthers {.}}{%
{Montabone}%
\ \protect \BOthers {.}}{%
{\protect \APACyear {2019, submitted to this issue}}%
}]{%
montabone2019}
\APACinsertmetastar {%
montabone2019}%
\begin{APACrefauthors}%
{Montabone}, L.%
, {Spiga}, L.%
, {Kass}, D.%
, {Kleinb{\"o}hl}, A.%
, {Forget}, F.%
\BCBL {}\ \BBA {} {Ehouarn}, M.%
\end{APACrefauthors}%
\unskip\
\newblock
\APACrefYearMonthDay{2019, submitted to this issue}{}{}.
\newblock
{\BBOQ}\APACrefatitle {{Martian Year 34 Column Dust Climatology from {M}ars
  {C}limate {S}ounder Observations: {R}econstructed Maps and Model
  Simulations}} {{Martian Year 34 Column Dust Climatology from {M}ars {C}limate
  {S}ounder Observations: {R}econstructed Maps and Model Simulations}}.{\BBCQ}
\newblock
\APACjournalVolNumPages{Submitted to J. Geophys. Res.}{}{}{}.
\newblock
\begin{APACrefDOI} \doi{arXiv:1907.08187} \end{APACrefDOI}
\PrintBackRefs{\CurrentBib}

\bibitem [\protect \citeauthoryear {%
{Morgan}%
\ \protect \BOthers {.}}{%
{Morgan}%
\ \protect \BOthers {.}}{%
{\protect \APACyear {2008}}%
}]{%
morgan2008}
\APACinsertmetastar {%
morgan2008}%
\begin{APACrefauthors}%
{Morgan}, D\BPBI D.%
, {Gurnett}, D\BPBI A.%
, {Kirchner}, D\BPBI L.%
, {Fox}, J\BPBI L.%
, {Nielsen}, E.%
\BCBL {}\ \BBA {} {Plaut}, J\BPBI J.%
\end{APACrefauthors}%
\unskip\
\newblock
\APACrefYearMonthDay{2008}{}{}.
\newblock
{\BBOQ}\APACrefatitle {Variation of the martian ionospheric electron density
  from {M}ars {E}xpress radar soundings} {Variation of the martian ionospheric
  electron density from {M}ars {E}xpress radar soundings}.{\BBCQ}
\newblock
\APACjournalVolNumPages{J. Geophys. Res.}{113}{}{A09303, 10.1029/2008JA013313}.
\newblock
\begin{APACrefDOI} \doi{10.1029/2008JA013313} \end{APACrefDOI}
\PrintBackRefs{\CurrentBib}

\bibitem [\protect \citeauthoryear {%
{N{\v{e}}mec}%
, {Morgan}%
\BCBL {}\ \BBA {} {Gurnett}%
}{%
{N{\v{e}}mec}%
\ \protect \BOthers {.}}{%
{\protect \APACyear {2016}}%
}]{%
nemec2016}
\APACinsertmetastar {%
nemec2016}%
\begin{APACrefauthors}%
{N{\v{e}}mec}, F.%
, {Morgan}, D\BPBI D.%
\BCBL {}\ \BBA {} {Gurnett}, D\BPBI A.%
\end{APACrefauthors}%
\unskip\
\newblock
\APACrefYearMonthDay{2016}{}{}.
\newblock
{\BBOQ}\APACrefatitle {{On improving the accuracy of electron density profiles
  obtained at high altitudes by the ionospheric sounder on the {M}ars {E}xpress
  spacecraft}} {{On improving the accuracy of electron density profiles
  obtained at high altitudes by the ionospheric sounder on the {M}ars {E}xpress
  spacecraft}}.{\BBCQ}
\newblock
\APACjournalVolNumPages{J. Geophys. Res.}{121}{10}{10,117-10,129}.
\newblock
\begin{APACrefDOI} \doi{10.1002/2016JA023054} \end{APACrefDOI}
\PrintBackRefs{\CurrentBib}

\bibitem [\protect \citeauthoryear {%
{Picardi}%
\ \protect \BOthers {.}}{%
{Picardi}%
\ \protect \BOthers {.}}{%
{\protect \APACyear {2004}}%
}]{%
picardi2004}
\APACinsertmetastar {%
picardi2004}%
\begin{APACrefauthors}%
{Picardi}, G.%
, {Biccari}, D.%
, {Seu}, R.%
, {Plaut}, J.%
, {Johnson}, W\BPBI T\BPBI K.%
, {Jordan}, R\BPBI L.%
\BDBL {}{Zampolini}, E.%
\end{APACrefauthors}%
\unskip\
\newblock
\APACrefYearMonthDay{2004}{}{}.
\newblock
{\BBOQ}\APACrefatitle {{MARSIS}: {M}ars {A}dvanced {R}adar for {S}ubsurface and
  {I}onosphere {S}ounding} {{MARSIS}: {M}ars {A}dvanced {R}adar for
  {S}ubsurface and {I}onosphere {S}ounding}.{\BBCQ}
\newblock
\BIn{} (\BPG~51-69).
\newblock
\APACaddressPublisher{}{ESA SP-1240: Mars Express: the Scientific Payload,
  available online at
  http://sci.esa.int/science-e/www/object/index.cfm?fobjectid=34885}.
\PrintBackRefs{\CurrentBib}

\bibitem [\protect \citeauthoryear {%
{Qin}%
\ \protect \BOthers {.}}{%
{Qin}%
\ \protect \BOthers {.}}{%
{\protect \APACyear {2019}}%
}]{%
qin2019}
\APACinsertmetastar {%
qin2019}%
\begin{APACrefauthors}%
{Qin}, J\BPBI F.%
, {Zou}, H.%
, {Ye}, Y\BPBI G.%
, {Yin}, Z\BPBI F.%
, {Wang}, J\BPBI S.%
\BCBL {}\ \BBA {} {Nielsen}, E.%
\end{APACrefauthors}%
\unskip\
\newblock
\APACrefYearMonthDay{2019}{}{}.
\newblock
{\BBOQ}\APACrefatitle {{Effects of Local Dust Storms on the Upper Atmosphere of
  {M}ars: {O}bservations and Simulations}} {{Effects of Local Dust Storms on
  the Upper Atmosphere of {M}ars: {O}bservations and Simulations}}.{\BBCQ}
\newblock
\APACjournalVolNumPages{J. Geophys. Res.}{124}{2}{602-616}.
\newblock
\begin{APACrefDOI} \doi{10.1029/2018JE005864} \end{APACrefDOI}
\PrintBackRefs{\CurrentBib}

\bibitem [\protect \citeauthoryear {%
S{\'a}nchez-Lavega%
, del Río-Gaztelurrutia%
, Hernández-Bernal%
\BCBL {}\ \BBA {} Delcroix%
}{%
S{\'a}nchez-Lavega%
\ \protect \BOthers {.}}{%
{\protect \APACyear {2019}}%
}]{%
sanchez2019}
\APACinsertmetastar {%
sanchez2019}%
\begin{APACrefauthors}%
S{\'a}nchez-Lavega, A.%
, del Río-Gaztelurrutia, T.%
, Hernández-Bernal, J.%
\BCBL {}\ \BBA {} Delcroix, M.%
\end{APACrefauthors}%
\unskip\
\newblock
\APACrefYearMonthDay{2019}{}{}.
\newblock
{\BBOQ}\APACrefatitle {The Onset and Growth of the 2018 {M}artian Global Dust
  Storm} {The onset and growth of the 2018 {M}artian global dust storm}.{\BBCQ}
\newblock
\APACjournalVolNumPages{Geophys Res. Lett.}{46}{11}{6101-6108}.
\newblock
\begin{APACrefURL}
  \url{https://agupubs.onlinelibrary.wiley.com/doi/abs/10.1029/2019GL083207}
  \end{APACrefURL}
\newblock
\begin{APACrefDOI} \doi{10.1029/2019GL083207} \end{APACrefDOI}
\PrintBackRefs{\CurrentBib}

\bibitem [\protect \citeauthoryear {%
Schunk%
\ \BBA {} Nagy%
}{%
Schunk%
\ \BBA {} Nagy%
}{%
{\protect \APACyear {2009}}%
}]{%
schunk2009}
\APACinsertmetastar {%
schunk2009}%
\begin{APACrefauthors}%
Schunk, R\BPBI W.%
\BCBT {}\ \BBA {} Nagy, A\BPBI F.%
\end{APACrefauthors}%
\unskip\
\newblock
\APACrefYear{2009}.
\newblock
\APACrefbtitle {Ionospheres} {Ionospheres}\ (\PrintOrdinal{Second}\ \BEd).
\newblock
\APACaddressPublisher{New York}{Cambridge University Press}.
\PrintBackRefs{\CurrentBib}

\bibitem [\protect \citeauthoryear {%
{Siddle}%
, {Mueller-Wodarg}%
, {Stone}%
\BCBL {}\ \BBA {} {Yelle}%
}{%
{Siddle}%
\ \protect \BOthers {.}}{%
{\protect \APACyear {2019}}%
}]{%
siddle2019}
\APACinsertmetastar {%
siddle2019}%
\begin{APACrefauthors}%
{Siddle}, A\BPBI G.%
, {Mueller-Wodarg}, I\BPBI C\BPBI F.%
, {Stone}, S\BPBI W.%
\BCBL {}\ \BBA {} {Yelle}, R\BPBI V.%
\end{APACrefauthors}%
\unskip\
\newblock
\APACrefYearMonthDay{2019}{}{}.
\newblock
{\BBOQ}\APACrefatitle {{Global characteristics of gravity waves in the upper
  atmosphere of {M}ars as measured by {MAVEN/NGIMS}}} {{Global characteristics
  of gravity waves in the upper atmosphere of {M}ars as measured by
  {MAVEN/NGIMS}}}.{\BBCQ}
\newblock
\APACjournalVolNumPages{Icarus}{333}{}{12-21}.
\newblock
\begin{APACrefDOI} \doi{10.1016/j.icarus.2019.05.021} \end{APACrefDOI}
\PrintBackRefs{\CurrentBib}

\bibitem [\protect \citeauthoryear {%
{Terada}%
\ \protect \BOthers {.}}{%
{Terada}%
\ \protect \BOthers {.}}{%
{\protect \APACyear {2017}}%
}]{%
terada2017}
\APACinsertmetastar {%
terada2017}%
\begin{APACrefauthors}%
{Terada}, N.%
, {Leblanc}, F.%
, {Nakagawa}, H.%
, {Medvedev}, A\BPBI S.%
, {Yi{\v{g}}it}, E.%
, {Kuroda}, T.%
\BDBL {}{Jakosky}, B\BPBI M.%
\end{APACrefauthors}%
\unskip\
\newblock
\APACrefYearMonthDay{2017}{}{}.
\newblock
{\BBOQ}\APACrefatitle {{Global distribution and parameter dependences of
  gravity wave activity in the Martian upper thermosphere derived from
  {MAVEN/NGIMS} observations}} {{Global distribution and parameter dependences
  of gravity wave activity in the Martian upper thermosphere derived from
  {MAVEN/NGIMS} observations}}.{\BBCQ}
\newblock
\APACjournalVolNumPages{J. Geophys. Res.}{122}{2}{2374-2397}.
\newblock
\begin{APACrefDOI} \doi{10.1002/2016JA023476} \end{APACrefDOI}
\PrintBackRefs{\CurrentBib}

\bibitem [\protect \citeauthoryear {%
{Thiemann}%
\ \protect \BOthers {.}}{%
{Thiemann}%
\ \protect \BOthers {.}}{%
{\protect \APACyear {2017}}%
}]{%
thiemann2017}
\APACinsertmetastar {%
thiemann2017}%
\begin{APACrefauthors}%
{Thiemann}, E\BPBI M\BPBI B.%
, {Chamberlin}, P\BPBI C.%
, {Eparvier}, F\BPBI G.%
, {Templeman}, B.%
, {Woods}, T\BPBI N.%
, {Bougher}, S\BPBI W.%
\BCBL {}\ \BBA {} {Jakosky}, B\BPBI M.%
\end{APACrefauthors}%
\unskip\
\newblock
\APACrefYearMonthDay{2017}{}{}.
\newblock
{\BBOQ}\APACrefatitle {{The {MAVEN EUVM} model of solar spectral irradiance
  variability at {M}ars: {A}lgorithms and results}} {{The {MAVEN EUVM} model of
  solar spectral irradiance variability at {M}ars: {A}lgorithms and
  results}}.{\BBCQ}
\newblock
\APACjournalVolNumPages{J. Geophys. Res.}{122}{}{2748-2767}.
\newblock
\begin{APACrefDOI} \doi{10.1002/2016JA023512} \end{APACrefDOI}
\PrintBackRefs{\CurrentBib}

\bibitem [\protect \citeauthoryear {%
{Thiemann}%
\ \protect \BOthers {.}}{%
{Thiemann}%
\ \protect \BOthers {.}}{%
{\protect \APACyear {2018}}%
}]{%
thiemann2018a}
\APACinsertmetastar {%
thiemann2018a}%
\begin{APACrefauthors}%
{Thiemann}, E\BPBI M\BPBI B.%
, {Eparvier}, F\BPBI G.%
, {Bougher}, S\BPBI W.%
, {Dominique}, M.%
, {Andersson}, L.%
, {Girazian}, Z.%
\BDBL {}{Jakosky}, B\BPBI M.%
\end{APACrefauthors}%
\unskip\
\newblock
\APACrefYearMonthDay{2018}{}{}.
\newblock
{\BBOQ}\APACrefatitle {{Mars Thermospheric Variability Revealed by {MAVEN EUVM}
  Solar Occultations: {S}tructure at Aphelion and Perihelion and Response to
  {EUV} Forcing}} {{Mars Thermospheric Variability Revealed by {MAVEN EUVM}
  Solar Occultations: {S}tructure at Aphelion and Perihelion and Response to
  {EUV} Forcing}}.{\BBCQ}
\newblock
\APACjournalVolNumPages{J. Geophys. Res.}{123}{9}{2248-2269}.
\newblock
\begin{APACrefDOI} \doi{10.1029/2018JE005550} \end{APACrefDOI}
\PrintBackRefs{\CurrentBib}

\bibitem [\protect \citeauthoryear {%
{Vandaele}%
\ \protect \BOthers {.}}{%
{Vandaele}%
\ \protect \BOthers {.}}{%
{\protect \APACyear {2019}}%
}]{%
vandaele2019}
\APACinsertmetastar {%
vandaele2019}%
\begin{APACrefauthors}%
{Vandaele}, A\BPBI C.%
, {Korablev}, O.%
, {Daerden}, F.%
, {Aoki}, S.%
, {Thomas}, I\BPBI R.%
, {Altieri}, F.%
\BDBL {}{ACS Science Team}%
\end{APACrefauthors}%
\unskip\
\newblock
\APACrefYearMonthDay{2019}{}{}.
\newblock
{\BBOQ}\APACrefatitle {{Martian dust storm impact on atmospheric {H}$_{2}$O and
  {D/H} observed by {E}xo{M}ars {T}race {G}as {O}rbiter}} {{Martian dust storm
  impact on atmospheric {H}$_{2}$O and {D/H} observed by {E}xo{M}ars {T}race
  {G}as {O}rbiter}}.{\BBCQ}
\newblock
\APACjournalVolNumPages{Nature}{568}{7753}{521-525}.
\newblock
\begin{APACrefDOI} \doi{10.1038/s41586-019-1097-3} \end{APACrefDOI}
\PrintBackRefs{\CurrentBib}

\bibitem [\protect \citeauthoryear {%
{Vogt}%
\ \protect \BOthers {.}}{%
{Vogt}%
\ \protect \BOthers {.}}{%
{\protect \APACyear {2017}}%
}]{%
vogt2017}
\APACinsertmetastar {%
vogt2017}%
\begin{APACrefauthors}%
{Vogt}, M\BPBI F.%
, {Withers}, P.%
, {Fallows}, K.%
, {Andersson}, L.%
, {Girazian}, Z.%
, {Mahaffy}, P\BPBI R.%
\BDBL {}{Jakosky}, B\BPBI M.%
\end{APACrefauthors}%
\unskip\
\newblock
\APACrefYearMonthDay{2017}{}{}.
\newblock
{\BBOQ}\APACrefatitle {{{MAVEN} observations of dayside peak electron densities
  in the ionosphere of {M}ars}} {{{MAVEN} observations of dayside peak electron
  densities in the ionosphere of {M}ars}}.{\BBCQ}
\newblock
\APACjournalVolNumPages{J. Geophys. Res.}{122}{}{891-906}.
\newblock
\begin{APACrefDOI} \doi{10.1002/2016JA023473} \end{APACrefDOI}
\PrintBackRefs{\CurrentBib}

\bibitem [\protect \citeauthoryear {%
{Wang}%
\ \BBA {} {Nielsen}%
}{%
{Wang}%
\ \BBA {} {Nielsen}%
}{%
{\protect \APACyear {2003}}%
}]{%
wang2003b}
\APACinsertmetastar {%
wang2003b}%
\begin{APACrefauthors}%
{Wang}, J\BHBI S.%
\BCBT {}\ \BBA {} {Nielsen}, E.%
\end{APACrefauthors}%
\unskip\
\newblock
\APACrefYearMonthDay{2003}{}{}.
\newblock
{\BBOQ}\APACrefatitle {{Behavior of the {M}artian dayside electron density peak
  during global dust storms}} {{Behavior of the {M}artian dayside electron
  density peak during global dust storms}}.{\BBCQ}
\newblock
\APACjournalVolNumPages{Planet. Space Sci.}{51}{}{329-338}.
\newblock
\begin{APACrefDOI} \doi{10.1016/S0032-0633(03)00015-1} \end{APACrefDOI}
\PrintBackRefs{\CurrentBib}

\bibitem [\protect \citeauthoryear {%
{Withers}%
}{%
{Withers}%
}{%
{\protect \APACyear {2006}}%
}]{%
withers2006a}
\APACinsertmetastar {%
withers2006a}%
\begin{APACrefauthors}%
{Withers}, P.%
\end{APACrefauthors}%
\unskip\
\newblock
\APACrefYearMonthDay{2006}{}{}.
\newblock
{\BBOQ}\APACrefatitle {{M}ars {G}lobal {S}urveyor and {M}ars {O}dyssey
  Accelerometer observations of the martian upper atmosphere during
  aerobraking} {{M}ars {G}lobal {S}urveyor and {M}ars {O}dyssey accelerometer
  observations of the martian upper atmosphere during aerobraking}.{\BBCQ}
\newblock
\APACjournalVolNumPages{Geophys. Res. Lett.}{33}{}{L02201,
  10.1029/2005GL024447}.
\newblock
\begin{APACrefDOI} \doi{10.1029/2005GL024447} \end{APACrefDOI}
\PrintBackRefs{\CurrentBib}

\bibitem [\protect \citeauthoryear {%
{Withers}%
}{%
{Withers}%
}{%
{\protect \APACyear {2009}}%
}]{%
withers2009a}
\APACinsertmetastar {%
withers2009a}%
\begin{APACrefauthors}%
{Withers}, P.%
\end{APACrefauthors}%
\unskip\
\newblock
\APACrefYearMonthDay{2009}{}{}.
\newblock
{\BBOQ}\APACrefatitle {A review of observed variability in the dayside
  ionosphere of {M}ars} {A review of observed variability in the dayside
  ionosphere of {M}ars}.{\BBCQ}
\newblock
\APACjournalVolNumPages{Adv. Space Res.}{44}{}{277-307}.
\newblock
\begin{APACrefDOI} \doi{10.1016/j.asr.2009.04.027} \end{APACrefDOI}
\PrintBackRefs{\CurrentBib}

\bibitem [\protect \citeauthoryear {%
{Withers}%
\ \protect \BOthers {.}}{%
{Withers}%
\ \protect \BOthers {.}}{%
{\protect \APACyear {2018}}%
}]{%
withers2018}
\APACinsertmetastar {%
withers2018}%
\begin{APACrefauthors}%
{Withers}, P.%
, {Felici}, M.%
, {Mendillo}, M.%
, {Moore}, L.%
, {Narvaez}, C.%
, {Vogt}, M\BPBI F.%
\BCBL {}\ \BBA {} {Jakosky}, B\BPBI M.%
\end{APACrefauthors}%
\unskip\
\newblock
\APACrefYearMonthDay{2018}{}{}.
\newblock
{\BBOQ}\APACrefatitle {{First {I}onospheric {R}esults {F}rom the {MAVEN}
  {R}adio {O}ccultation {S}cience {E}xperiment ({ROSE})}} {{First {I}onospheric
  {R}esults {F}rom the {MAVEN} {R}adio {O}ccultation {S}cience {E}xperiment
  ({ROSE})}}.{\BBCQ}
\newblock
\APACjournalVolNumPages{J. Geophys. Res.}{123}{}{4171-4180}.
\newblock
\begin{APACrefDOI} \doi{10.1029/2018JA025182} \end{APACrefDOI}
\PrintBackRefs{\CurrentBib}

\bibitem [\protect \citeauthoryear {%
{Withers}%
\ \BBA {} {Pratt}%
}{%
{Withers}%
\ \BBA {} {Pratt}%
}{%
{\protect \APACyear {2013}}%
}]{%
withers2013a}
\APACinsertmetastar {%
withers2013a}%
\begin{APACrefauthors}%
{Withers}, P.%
\BCBT {}\ \BBA {} {Pratt}, R.%
\end{APACrefauthors}%
\unskip\
\newblock
\APACrefYearMonthDay{2013}{}{}.
\newblock
{\BBOQ}\APACrefatitle {{An observational study of the response of the upper
  atmosphere of Mars to lower atmospheric dust storms}} {{An observational
  study of the response of the upper atmosphere of Mars to lower atmospheric
  dust storms}}.{\BBCQ}
\newblock
\APACjournalVolNumPages{Icarus}{225}{}{378-389}.
\newblock
\begin{APACrefDOI} \doi{10.1016/j.icarus.2013.02.032} \end{APACrefDOI}
\PrintBackRefs{\CurrentBib}

\bibitem [\protect \citeauthoryear {%
Wolkenberg%
\ \protect \BOthers {.}}{%
Wolkenberg%
\ \protect \BOthers {.}}{%
{\protect \APACyear {2018}}%
}]{%
wolkenberg2018}
\APACinsertmetastar {%
wolkenberg2018}%
\begin{APACrefauthors}%
Wolkenberg, P.%
, Giuranna, M.%
, Grassi, D.%
, Aronica, A.%
, Aoki, S.%
, Scaccabarozzi, D.%
\BCBL {}\ \BBA {} Saggin, B.%
\end{APACrefauthors}%
\unskip\
\newblock
\APACrefYearMonthDay{2018}{}{}.
\newblock
{\BBOQ}\APACrefatitle {Characterization of dust activity on {M}ars from {MY}27
  to {MY}32 by {PFS-MEX} observations} {Characterization of dust activity on
  {M}ars from {MY}27 to {MY}32 by {PFS-MEX} observations}.{\BBCQ}
\newblock
\APACjournalVolNumPages{Icarus}{310}{}{32 - 47}.
\newblock
\begin{APACrefDOI} \doi{https://doi.org/10.1016/j.icarus.2017.10.045}
  \end{APACrefDOI}
\PrintBackRefs{\CurrentBib}

\bibitem [\protect \citeauthoryear {%
{Zou}%
, {Lillis}%
, {Wang}%
\BCBL {}\ \BBA {} {Nielsen}%
}{%
{Zou}%
\ \protect \BOthers {.}}{%
{\protect \APACyear {2011}}%
}]{%
zou2011}
\APACinsertmetastar {%
zou2011}%
\begin{APACrefauthors}%
{Zou}, H.%
, {Lillis}, R\BPBI J.%
, {Wang}, J\BPBI S.%
\BCBL {}\ \BBA {} {Nielsen}, E.%
\end{APACrefauthors}%
\unskip\
\newblock
\APACrefYearMonthDay{2011}{}{}.
\newblock
{\BBOQ}\APACrefatitle {{Determination of seasonal variations in the {M}artian
  neutral atmosphere from observations of ionospheric peak height}}
  {{Determination of seasonal variations in the {M}artian neutral atmosphere
  from observations of ionospheric peak height}}.{\BBCQ}
\newblock
\APACjournalVolNumPages{J. Geophys. Res.}{116}{}{E09004}.
\newblock
\begin{APACrefDOI} \doi{10.1029/2011JE003833} \end{APACrefDOI}
\PrintBackRefs{\CurrentBib}

\bibitem [\protect \citeauthoryear {%
{Zou}%
\ \protect \BOthers {.}}{%
{Zou}%
\ \protect \BOthers {.}}{%
{\protect \APACyear {2016}}%
}]{%
zou2016}
\APACinsertmetastar {%
zou2016}%
\begin{APACrefauthors}%
{Zou}, H.%
, {Ye}, Y\BPBI G.%
, {Wang}, J\BPBI S.%
, {Nielsen}, E.%
, {Cui}, J.%
\BCBL {}\ \BBA {} {Wang}, X\BPBI D.%
\end{APACrefauthors}%
\unskip\
\newblock
\APACrefYearMonthDay{2016}{}{}.
\newblock
{\BBOQ}\APACrefatitle {{A method to estimate the neutral atmospheric density
  near the ionospheric main peak of {M}ars}} {{A method to estimate the neutral
  atmospheric density near the ionospheric main peak of {M}ars}}.{\BBCQ}
\newblock
\APACjournalVolNumPages{J. Geophys. Res.}{121}{}{3464-3475}.
\newblock
\begin{APACrefDOI} \doi{10.1002/2015JA022304} \end{APACrefDOI}
\PrintBackRefs{\CurrentBib}

\bibitem [\protect \citeauthoryear {%
{Zurek}%
\ \protect \BOthers {.}}{%
{Zurek}%
\ \protect \BOthers {.}}{%
{\protect \APACyear {2017}}%
}]{%
zurek2017}
\APACinsertmetastar {%
zurek2017}%
\begin{APACrefauthors}%
{Zurek}, R\BPBI W.%
, {Tolson}, R\BPBI A.%
, {Bougher}, S\BPBI W.%
, {Lugo}, R\BPBI A.%
, {Baird}, D\BPBI T.%
, {Bell}, J\BPBI M.%
\BCBL {}\ \BBA {} {Jakosky}, B\BPBI M.%
\end{APACrefauthors}%
\unskip\
\newblock
\APACrefYearMonthDay{2017}{}{}.
\newblock
{\BBOQ}\APACrefatitle {{Mars thermosphere as seen in {MAVEN} accelerometer
  data}} {{Mars thermosphere as seen in {MAVEN} accelerometer data}}.{\BBCQ}
\newblock
\APACjournalVolNumPages{J. Geophys. Res.}{122}{3}{3798-3814}.
\newblock
\begin{APACrefDOI} \doi{10.1002/2016JA023641} \end{APACrefDOI}
\PrintBackRefs{\CurrentBib}

\end{thebibliography}
%





\newpage
\begin{table}
 \caption{Summary of the observed ionospheric response to dust for the six Mars years (MY) analyzed in Section~\ref{sec:dustseasons}. 
 Columns 2-6 list the solar longitudes (L$_s$, $^{\circ}$), Earth years, dates (mm/dd), solar zenith angles (SZA,$^{\circ}$), and minimum and maximum latitudes (LAT, $^{\circ}$) of the MARSIS observations that were analyzed in each MY. Column 7 lists the EUV irradiance level (EUV, mW/m$^2$). Columns 8-9 summarize each dust storm, including the increase in the globally-averaged dust optical depth ($\Delta \tau$), and the maximum equatorial-averaged (between $\pm$30$^{\circ}$) dust optical depth ($\tau_{max}$). Columns 10-12 summarize the observed changes in the ionospheric peak, including its increase in altitude ($\Delta h_{max}$, km), and its variability before ($\sigma_{h_{max}}^i$, km) and after ($\sigma_{h_{max}}^f$, km) dust storm onset as defined throughout Section~\ref{sec:dustseasons}. }
 \centering
 \begin{tabular}{r r l l r r r r r r r r }
 MY                     & 
 L$_s$                  & 
 Year                   &
 Date                   &
 SZA                    &
 LAT               &
 EUV                    &
 $\Delta \tau$          & 
 $\tau_{max}$           &
 $\Delta h_{max}$       &
 $\sigma_{h_{max}}^i$   &
 $\sigma_{h_{max}}^f$ \vspace{-1mm} 
 \\ 
 \hline
   27  & 305-330  & 2005 & 10/12 - 11/24 & 20-50 & (-20, -70)   &  1.3  & 0.2 & 0.3 & 10-15 & 4 & 4\\
   28  & 250-295  & 2007 & 06/03 - 08/14 & 30-40 & (-50, -65)   &  1.6  & 0.6 & 1.2 & 10-15 & 4 & 11 \\
   29  & 230-245  & 2009 & 03/20 - 04/05 & 45-55 & (-65, -75)   &  1.2  & 0.1 & 0.5 & 10-20    & 11 & 11 \\
   32  & 215-230  & 2014 & 10/16 - 11/10 & 75-80 & (40, 50)   &  2.4  & 0.1 & 0.5 & 10-15 & 5 & 7 \\
   33  & 304-310  & 2017 & 01/23 - 02/21 & 70-80 & (-20, -40)   &  1.1  & 0.0 & 0.2 & $\sim$10 & -- & --\\
   34  & 180-195  & 2018 & 05/22 - 06/26 & 60-85 & (60, 85)   &  1.0  & 0.5 & 1.5& 10-15 & -- & -- \\
 \end{tabular}
 \end{table}
 \clearpage

\newpage

\newpage
\begin{figure}[ht!]
 \centering
\includegraphics[width=1\textwidth]{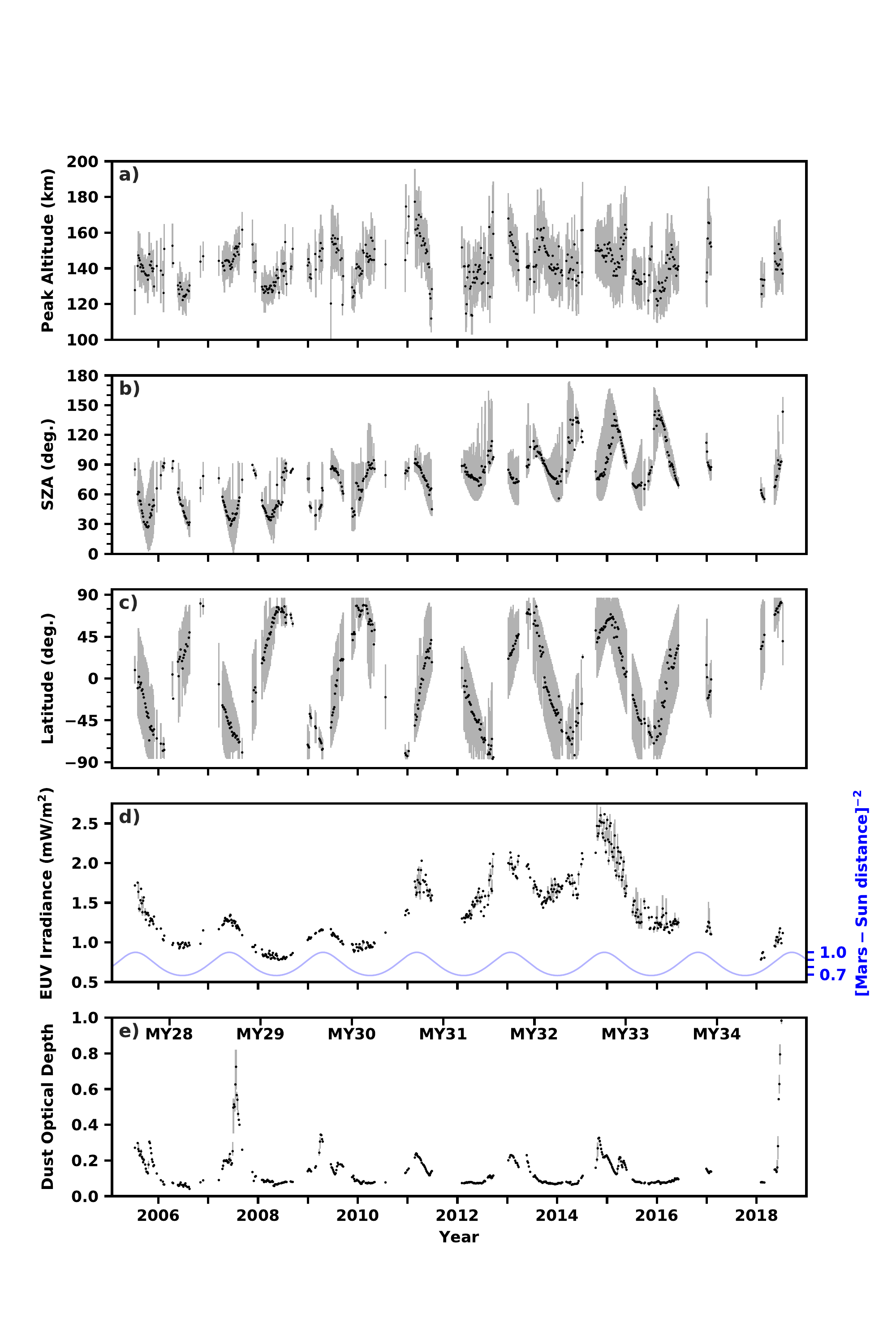}  
\caption{Time series of the data used in this study. Panels a-c show the ionospheric peak altitudes from MARSIS, their solar zenith angles (SZA), and their geographic latitudes. Panel d shows the solar EUV irradiance and the inverse-square of the Mars-Sun distance. Panel e shows the globally-averaged dust optical depths ($\tau$) during the MARSIS observations. The black circles in each panel are averages from 16 MEX orbits (120 hours). The error bars in Panel a, d, and e show the standard deviations from within each averaging bin, while the error bars in Panels b and c show the complete spread in the data within each averaging bin. 
}
\label{overview}
\end{figure}

\newpage    
\begin{figure}[ht!]
\centering
\includegraphics[width=0.7\textwidth]{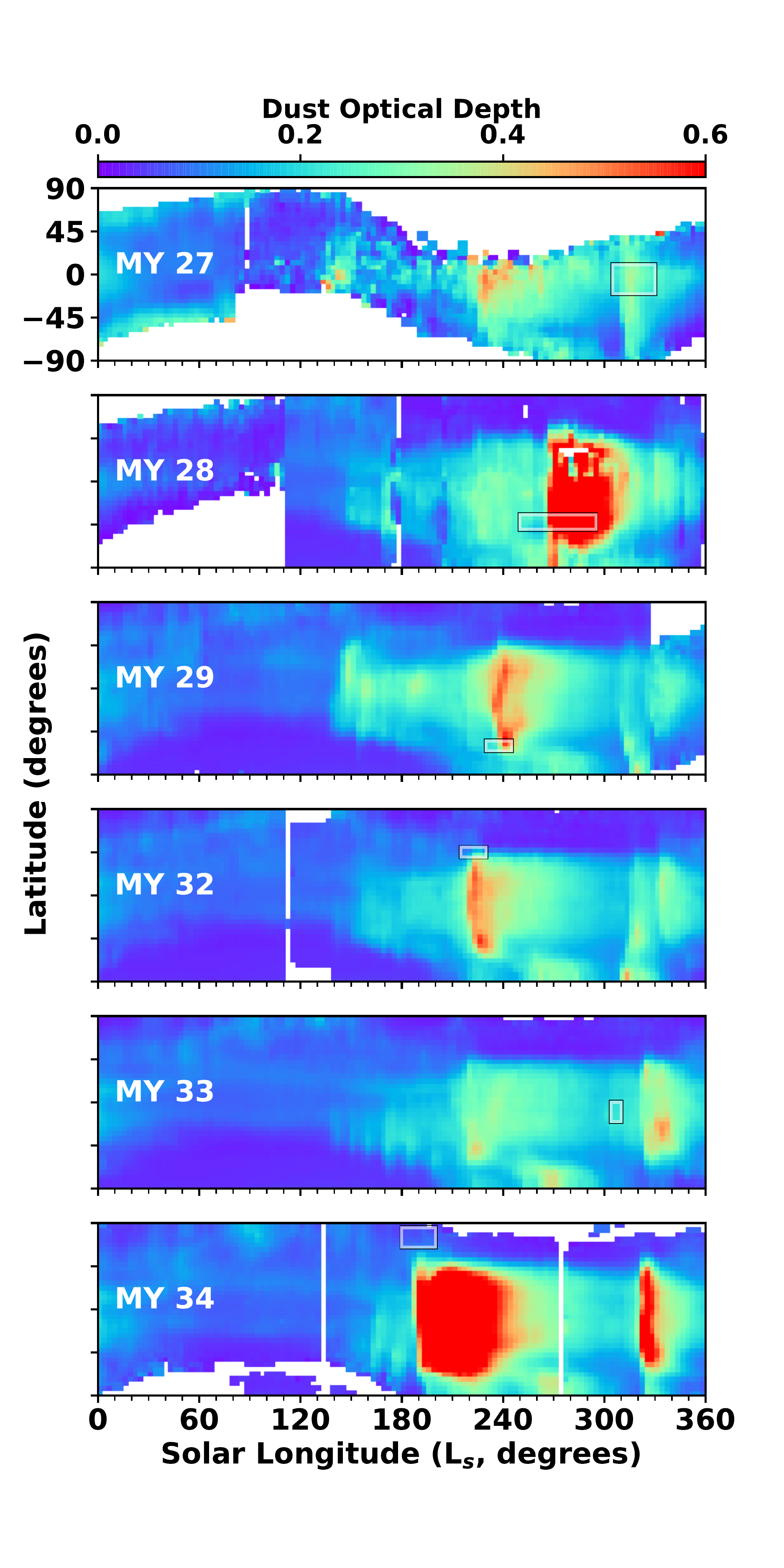}
\caption{Complete dust maps from the six Mars years during which we analyze MARSIS peak altitudes. The rectangles in each panel mark the L$_s$ and latitude coverage of the MARSIS observations. 
}
\label{dustmaps}
\end{figure}

\newpage    
\begin{figure}[ht!]
\centering
\includegraphics[width=0.7\textwidth]{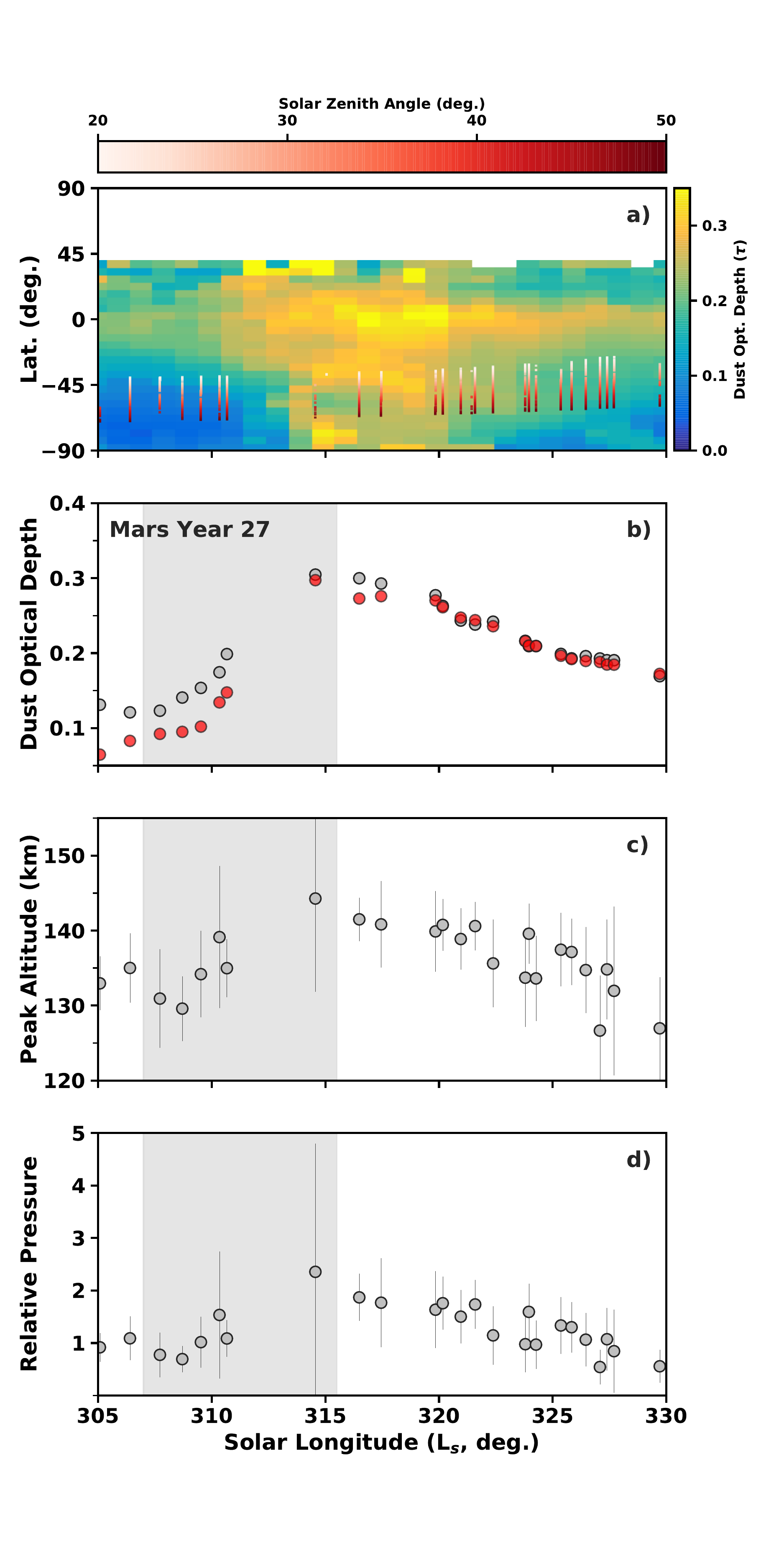}
\caption{The MY 27 dust season at L$_{s}$ 305$^{\circ}$ to 330$^{\circ}$. \textbf{a)} Dust optical depths averaged over 1$^{\circ}$ in L$_s$, 360$^{\circ}$ in longitude, and 5$^{\circ}$ in latitude. The MARSIS measurement coverage is plotted on top of the dust map, colored according to the SZA of the observation. Each vertical line represents the MARSIS latitudinal coverage from a single orbit. \textbf{b)} Global average (gray) and local average (red) dust optical depths from orbits during which there were at least 10 MARSIS peak altitude measurements. \textbf{c)} Orbit-averaged ionospheric peak altitudes. \textbf{d)} Orbit-averaged relative pressures derived using Equation~\ref{eq5} with a reference altitude of $h_{max_i}$ = 134 km. The gray shaded regions mark the period when the peak altitude and relative pressure increases during the dust storm.
}
\label{MY27}
\end{figure}

\newpage    
\begin{figure}[ht!]
\centering
\includegraphics[width=0.8\textwidth]{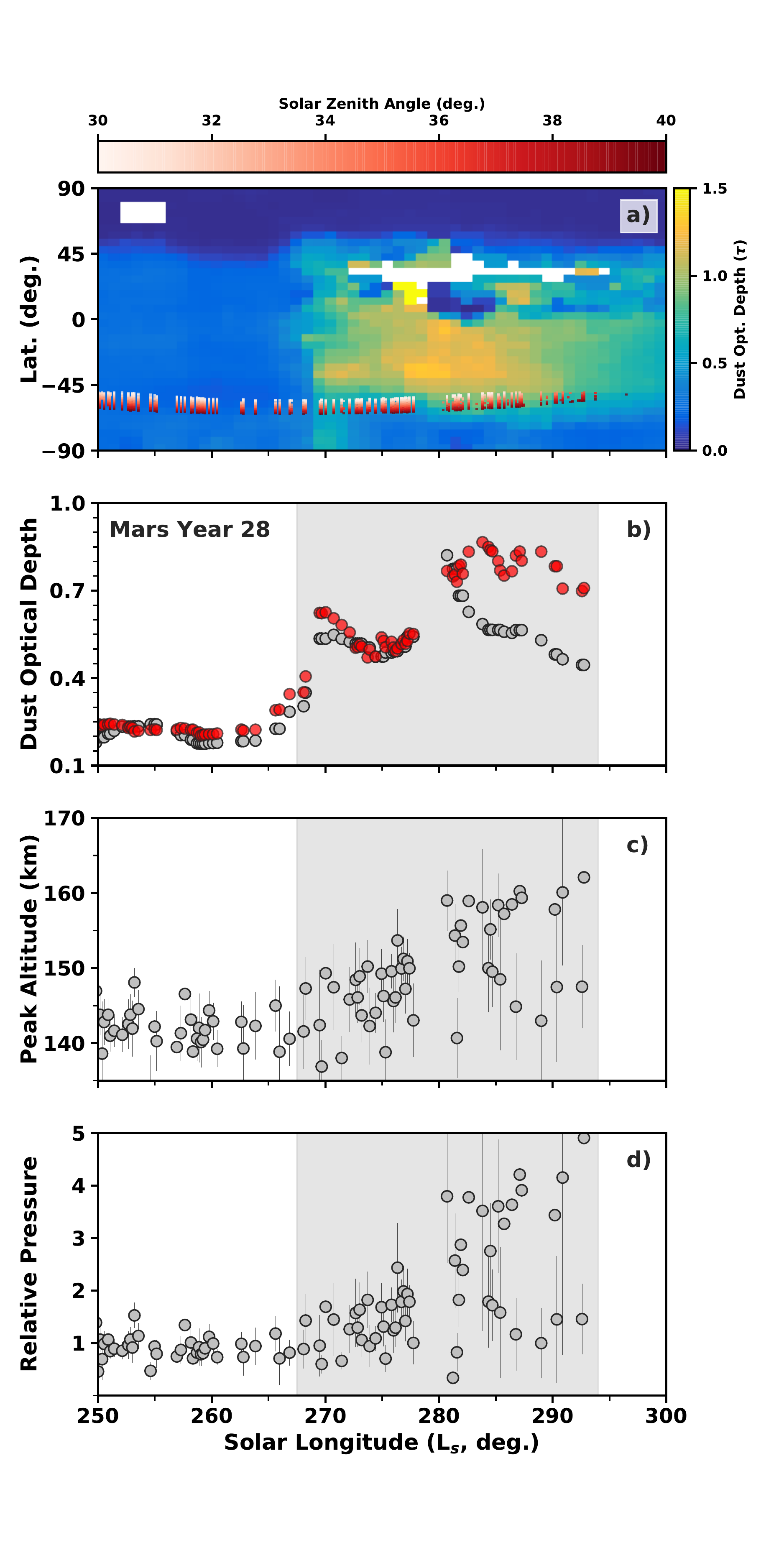}
\caption{Similar to Figure~\ref{MY27} but showing observations from the MY 28 global dust storm and with the axes scaled differently. The relative pressure is derived using a reference altitude of $h_{max_i}$ = 143 km.
}
\label{MY28}
\end{figure}

\newpage    
\begin{figure}[ht!]
\centering
\includegraphics[width=0.8\textwidth]{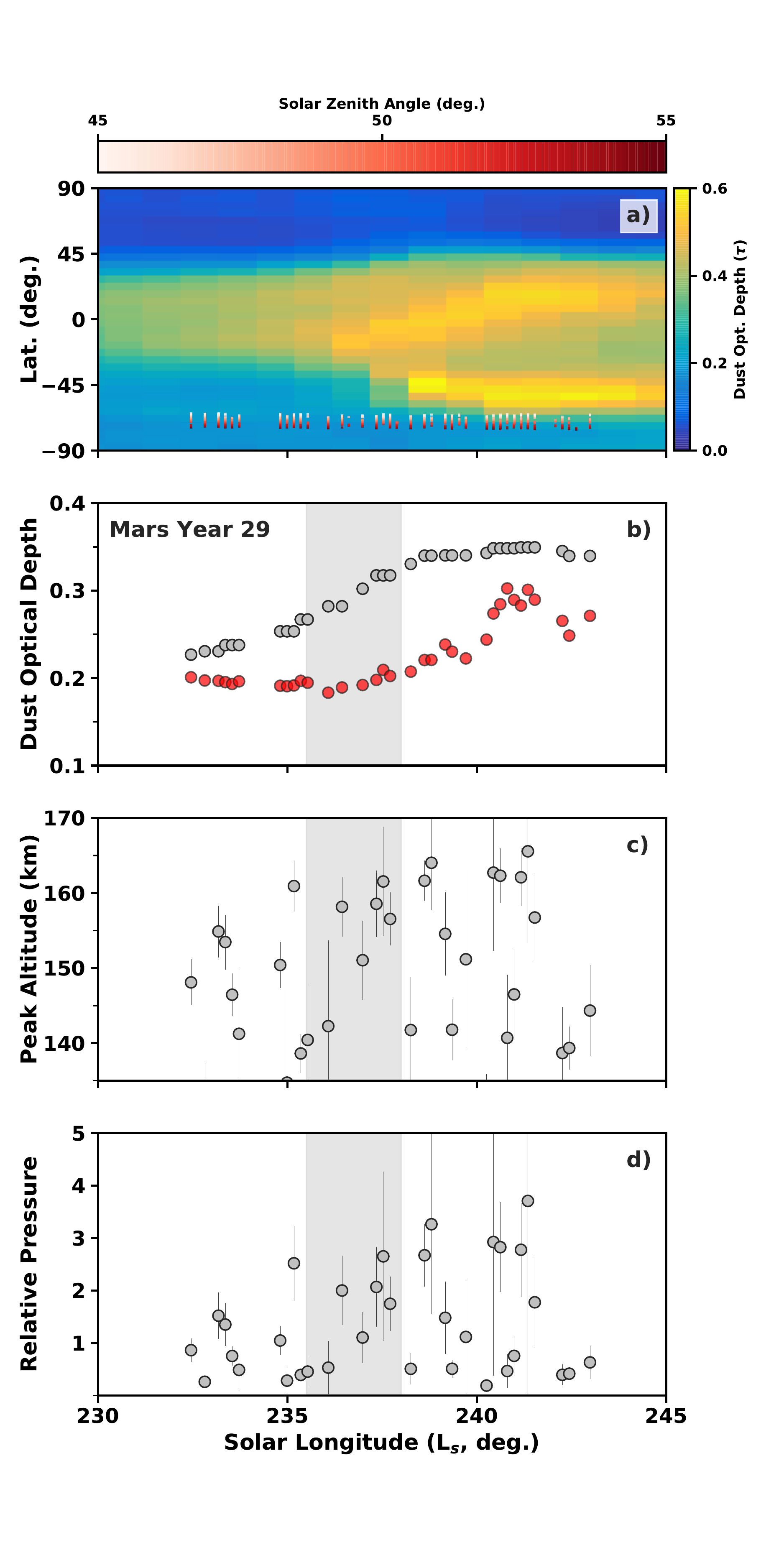}
\caption{Similar to Figure~\ref{MY27} but showing observations from the MY 29 dust season and with the axes scaled differently. The relative pressure is derived using a reference altitude of $h_{max_i}$ = 150 km.
}
\label{MY29}
\end{figure}

\newpage    
\begin{figure}[ht!]
\centering
\includegraphics[width=0.8\textwidth]{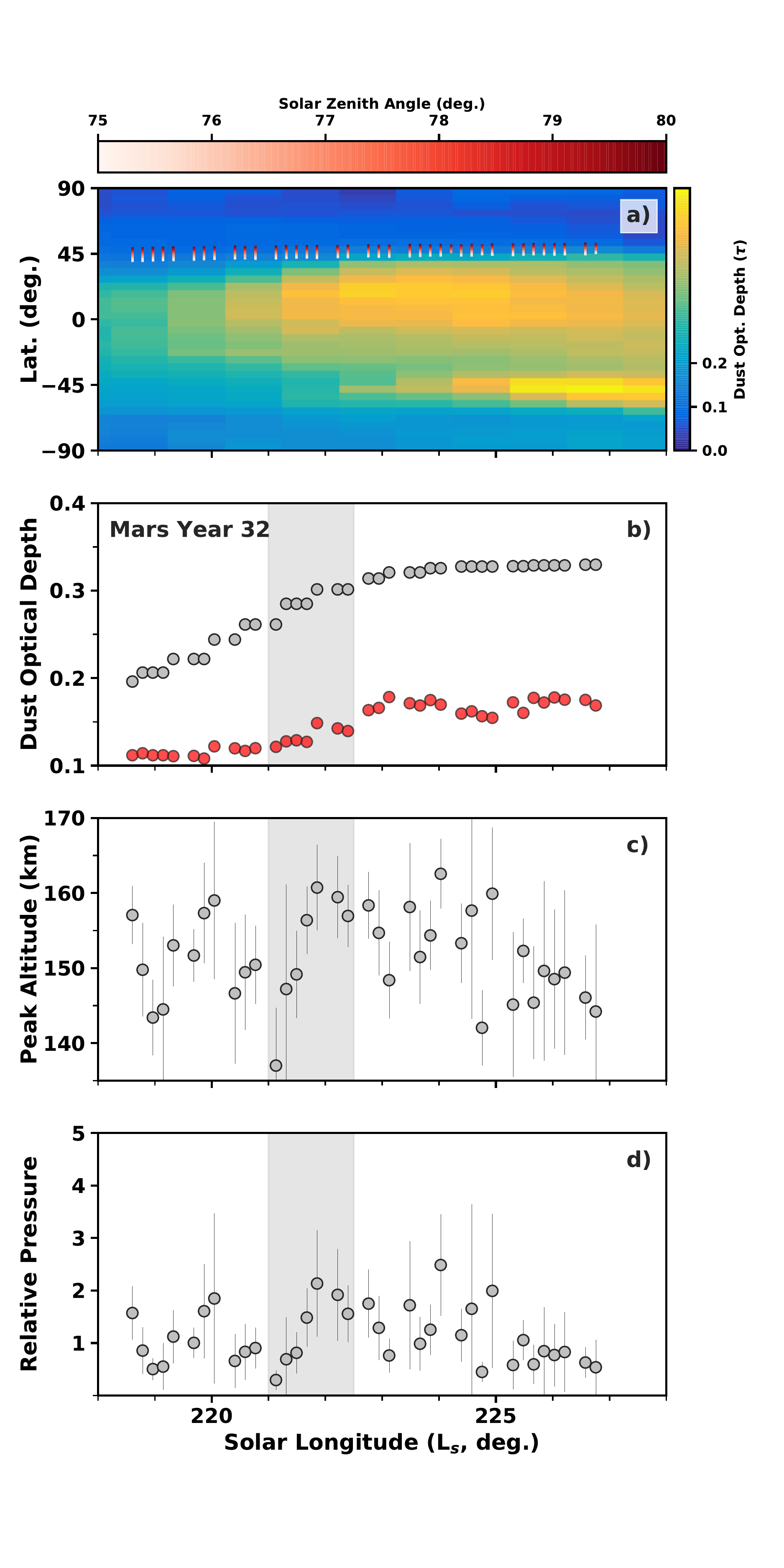}
\caption{Similar to Figure~\ref{MY27} but showing observations from the MY 32 dust season and with the axes scaled differently. The relative pressure is derived using a reference altitude of $h_{max_i}$ = 152 km.
}
\label{MY32}
\end{figure}

\newpage    
\begin{figure}[ht!]
\centering
\includegraphics[width=0.7\textwidth]{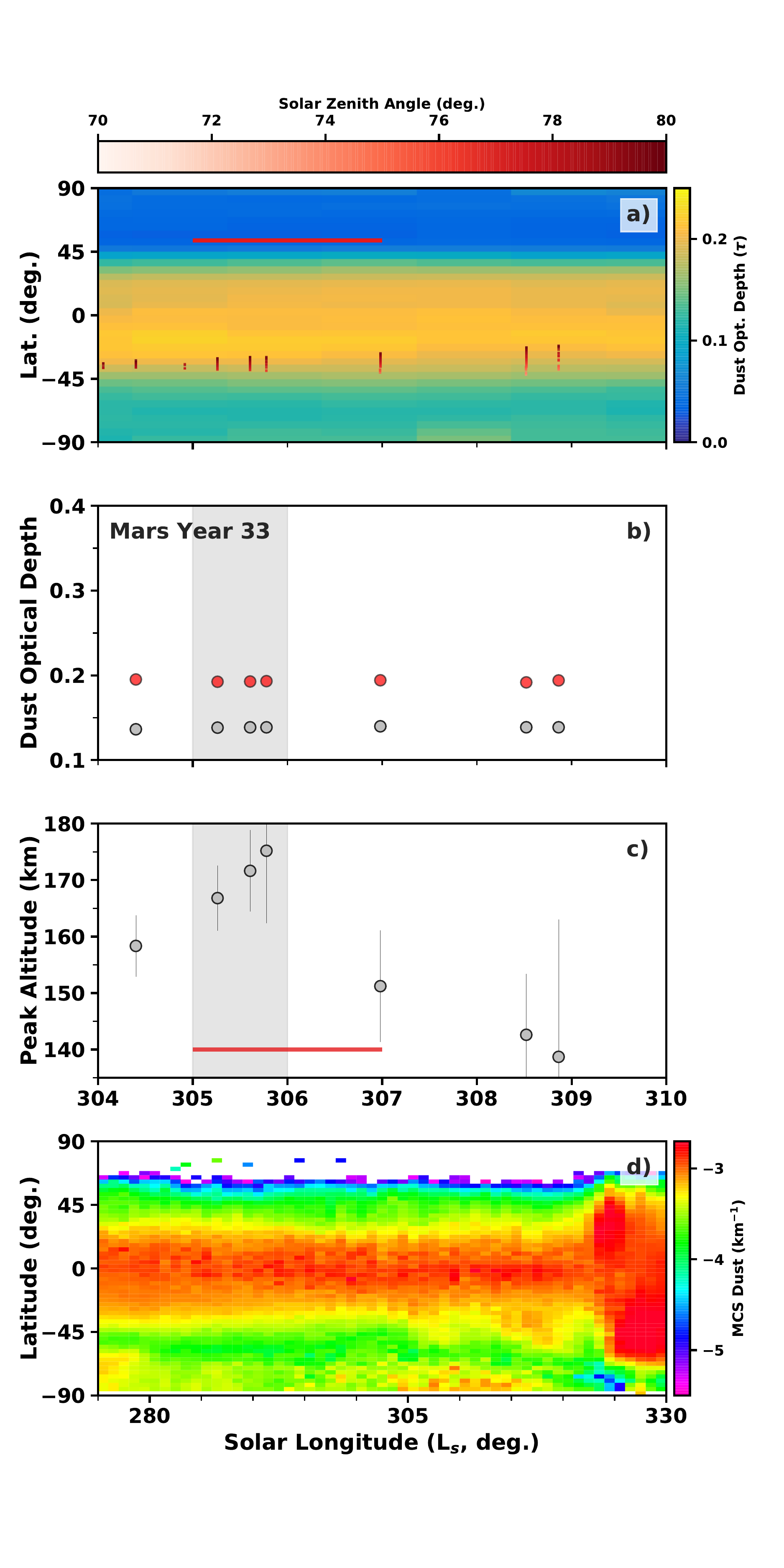}
\caption{Similar to Figure~\ref{MY27} but showing observations from MY 33 during the time when the MAVEN Radio Occultation Science Experiment (ROSE) observed the peak altitude rise during a localized dust storm \citep{withers2018}. The horizontal red lines mark the latitudes of the ROSE observations (Panel a) and the L$_s$ range when the peak was observed to rise (Panel c). Panel d shows a zoomed out map of the MCS dust extinction at 50 Pa ($\sim$25 km), which confirms the localized dust storm between L$_s$ $\sim$ 305$^{\circ}$-315$^{\circ}$ and at latitudes $<$ -60$^{\circ}$ that was reported in \citet{withers2018}.  }
\label{MY33}
\end{figure}

\newpage    
\begin{figure}[ht!]
\centering
\includegraphics[width=0.8\textwidth]{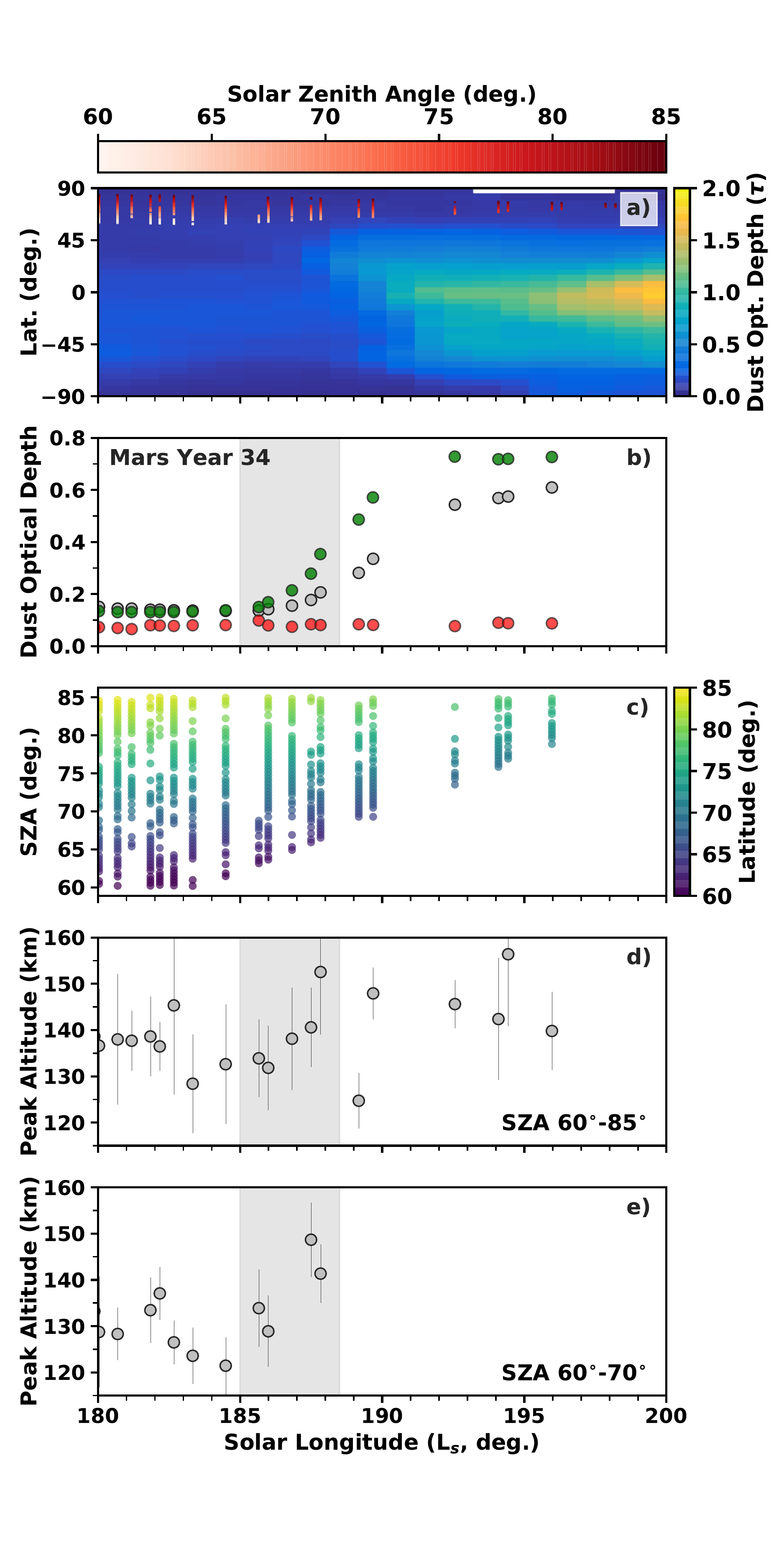}
\caption{Observations from the MY 34 global dust storm. Panels a and b are the same as in Figure~\ref{MY27}, but the northern hemisphere averaged dust optical depth (latitudes 0$^{\circ}$-45$^{\circ}$) was added to Panel b (green circles). \textbf{c)} The SZAs and latitudes of the MARSIS measurements. \textbf{d)} Orbit-averaged peak altitudes for SZAs 60$^{\circ}$-85$^{\circ}$. \textbf{e)} Orbit-averaged peak altitudes for SZAs 60$^{\circ}$-70$^{\circ}$.
}
\label{MY34}
\end{figure}

\end{document}